\documentclass[11pt,aps,preprintnumbers,nofootinbib,superscriptaddress,notitlepage,longbibliography]{revtex4-2}

\usepackage[T1]{fontenc}
\usepackage[english]{babel}
\usepackage{stix}
\usepackage{epsfig}
\usepackage{amsmath,bm,bbm,tensor,physics,dsfont}
\usepackage{graphicx}
\usepackage{verbatim}
\usepackage{csquotes}
\usepackage{simplewick}
\usepackage{appendix}
\usepackage[usenames,dvipsnames,svgnames]{xcolor}
\usepackage[normalem]{ulem}
\usepackage{hyperref}
\usepackage{cleveref}
\usepackage{orcidlink}

\definecolor{IAN}{RGB}{1,80,158}

\usepackage[framemethod=default]{mdframed}
\newmdenv[skipabove=7pt,
skipbelow=7pt,
rightline=false,
leftline=false,
topline=false,
bottomline=false,
backgroundcolor=gray!10,
linecolor=gray,
innerleftmargin=5pt,
innerrightmargin=5pt,
innertopmargin=5pt,
innerbottommargin=5pt,
leftmargin=0cm,
rightmargin=0cm,
linewidth=4pt]{eBox}

\newcommand{\md}{\mathrm{d}}
\newcommand\supsetsim{\mathrel{\substack{
  \textstyle\supset\\[-0.2ex]\textstyle\sim}}}

\begin{document}

\title{On the consistent disformal couplings to fermions}

\author{\textsc{Guillem Dom\`enech\,\orcidlink{0000-0003-2788-884X}}}
    \email{{guillem.domenech}@{itp.uni-hannover.de}}
\affiliation{Institute for Theoretical Physics, Leibniz University Hannover, Appelstraße 2, 30167 Hannover, Germany.}
\affiliation{ Max Planck Institute for Gravitational Physics, Albert Einstein Institute, 30167 Hannover, Germany.}

\author{\textsc{Alexander Ganz\,\orcidlink{0000-0001-7939-9058}}}
    \email{{alexander.ganz@itp.uni-hannover.de}}
    \affiliation{Institute for Theoretical Physics, Leibniz University Hannover, Appelstraße 2, 30167 Hannover, Germany.}

\author{\textsc{Apostolos Tsabodimos\,\orcidlink{0009-0001-0230-5647}}}
    \email{{apostolos.tsabodimos}@{stud.uni-hannover.de}}
    \affiliation{Institute for Theoretical Physics, Leibniz University Hannover, Appelstraße 2, 30167 Hannover, Germany.}
    \affiliation{Center for Physical Sciences and Technology, Sauletekio 3, 10257 Vilnius, Lithuania}

\begin{abstract}
  Disformal couplings to fermions lead to a unique derivative coupling to the axial fermionic current, which contains higher derivatives in general. We derive general conditions on consistent disformal couplings by requiring the absence of higher time derivatives, as they typically lead to ghost degrees of freedom. For a two-scalar field disformal transformation, we show that the consistent disformal coupling must have a degenerate field space metric. This allows us to explore consistent, new two-scalar field modified gravity models. We show that the transformation of the Einstein-Hilbert action leads to two-field Horndeski or two-field DHOST theories. Our formalism also applies to disformal transformations with higher derivatives. We derive the consistent subclasses of disformal transformations that include second derivatives of a scalar field and first derivatives of a vector field that lead to generalized U-DHOST and degenerate beyond generalized Proca theories.

\end{abstract}

\maketitle

\section{Introduction}

While General Relativity (GR) is successful in describing gravity on local scales, on cosmological scales, there remain open questions, most notably the nature of dark energy and dark matter. In that regard, it is plausible that the dark sector involves modifications of gravity \cite{Copeland:2006wr,Tsujikawa:2010zza,Gubitosi:2012hu,Gleyzes:2013ooa,Kobayashi:2019hrl,Frusciante:2019xia,Brax:2020gqg}. It is thus essential to explore general theories of gravity and their coupling to standard matter.

Among a wide variety of modified gravity models \cite{Tsujikawa:2010zza}, the most popular approaches are the inclusion of a scalar or vector field. In order to study these models in a unified manner, there has been a focus on constructing general scalar- or vector-tensor theories with second-order equations of motion. The former are known as Horndeski theories \cite{Horndeski:1974wa} ( rediscovered in the context of Galileon models~\cite{Charmousis:2011bf,Deffayet:2011gz}) while the latter as generalized Proca gravity \cite{Heisenberg:2014rta}. Since the equations of motion are second-order, there are no Ostrogradsky instabilities \cite{Ostrogradsky:1850fid}, commonly present in higher derivative theories. These would normally render the theory unstable, see, e.g., Refs.~\cite{Woodard:2015zca,Ganz:2020skf} and references therein. However, it was later realized that the restriction on the second order equation of motions can be relaxed by requiring that the kinetic matrix is degenerate, leading to Degenerate Higher Order Scalar-Tensor (DHOST) theories~\cite{Gleyzes:2014dya,Gleyzes:2014qga,Langlois:2015cwa,Langlois:2015skt,Crisostomi:2016czh,BenAchour:2016fzp} and beyond generalized Proca theories \cite{Heisenberg:2016eld,Kimura:2016rzw}. 

Within scalar-tensor and vector-tensor theories, disformal transformations \cite{auth:BekensteinOriginalDisformal} play a crucial role in constructing and exploring these models. In particular, it has been shown that disformal transformations involving the scalar field and its first derivatives leave the functional form of DHOST theories unchanged \cite{Bettoni:2013diz,Zumalacarregui:2013pma,Ezquiaga:2017ner,Crisostomi:2016czh,BenAchour:2016cay,BenAchour:2016fzp,deRham:2016wji}. This provides a useful method of classifying and mapping scalar-tensor theories. Similar work has been done for generalized Proca theories \cite{Domenech:2018vqj,Kimura:2016rzw}. 

With disformal transformations, one can remove part of the higher-derivative interactions in the gravitational sector and recast them as matter coupled to a new metric containing derivatives of the scalar field \cite{Brax:2012hm} or couplings to the vector field \cite{Ramazanoglu:2019jrr,Minamitsuji:2020pak}. In particular, there is a metric "frame" in which Gravitational Waves (GWs) propagate at the speed of light \cite{Fujita:2015ymn}.
Conversely, disformal couplings to the matter sector can be used to construct viable modified gravity models by performing the inverse disformal transformation. The number of degrees of freedom does not change \cite{Domenech:2015tca,Takahashi:2017zgr} if the metric transformation is invertible. Therefore, the resulting modified gravity model is free of Ostrogradsky ghosts, as long as there were no ghosts to begin with, that is, no ghosts due to the disformal couplings. 
In other words, investigating ghosts in higher derivative gravity theories with a minimal matter coupling is equivalent to looking for potential ghosts due to the disformal coupling. 

In a further generalization, Ref.~ \cite{auth:Takahashi_Invertible_Disformal,Takahashi:2022mew,Takahashi:2023vva,auth:Takahashi_Matter_Coupling,Naruko:2022vuh} included higher derivatives of the scalar field in the disformal transformation and used it to construct a class of \textit{disformal Horndeski}. This is a new type of modified gravity theory with higher derivatives beyond DHOST. However,  they showed that the couplings to fermions break 
the degeneracy condition and lead to ghosts. The absence of ghosts restricts the form of the disformal transformation \cite{auth:Takahashi_Matter_Coupling}.  Similar concepts of disformal transformation with higher derivatives have also been studied in \cite{Babichev:2021bim,Domenech:2019syf,Domenech:2023ryc,Babichev:2024eoh}.

Another direction for generalizations of scalar-tensor theories is multi-field extensions. These are often motivated by higher-dimensional compactification in string theory and have been extensively studied in the context of inflation \cite{Dimopoulos:2005ac,Kanti:1999vt,Kanti:1999ie,Silverstein:2003hf,Alishahiha:2004eh}. Like in the single field, one can consider multi-field derivative couplings to gravity. For example, Ref.~\cite{Padilla:2012dx} proposes a multi-field Galileon model, conjecturing that it is the most general Lagrangian leading to second-order equations of motion. However, it was later shown in Ref.~\cite{Kobayashi:2013ina} that these models are only a subset as they do not include multi-field DBI Galileon models \cite{Hinterbichler:2010xn,Renaux-Petel:2011rmu} (there are based on the unification of single DBI and Galileon models \cite{deRham:2010eu}). 

The construction of the most general multi-field DBI galielon models is expected to be linked to multi-field disformal transformations.
For two scalar fields, Ref.~\cite{Ohashi:2015fma} derived the most general equation of motion for two scalar fields with a second-order equation of motion. Ref.~\cite{Horndeski:2024hee} has later derived the corresponding Lagrangian, which we will call two-field Horndeski. We will explore here general consistent two-field disformal transformations.

In that respect, we have recently derived the general form of disformal couplings to Dirac fermions in Ref.~\cite{auth:us}. There, we were more concerned with constraints due to apparent Lorentz-violating effects. But we also demonstrated that disformal couplings generically lead to derivative coupling between the disformal fields and the fermionic axial current. In this paper, we extend our previous work by investigating under which conditions such a coupling does not lead to ghost degrees of freedom. 
Such consistent transformations can be used to construct new higher derivative theories of gravity. We will demonstrate the potential with some simpler cases.

The structure of the paper is as follows. In section \ref{sec:Preliminaries}, we review the disformal coupling to fermions as outlined in \cite{auth:us} and discuss the requirements for a consistent disformal matter coupling. In section \ref{sec:Two_scalar_fields}, we explore the disformal coupling with two scalar fields, showing that degeneracy conditions are needed for consistency, which leads to a degenerate field space metric. We briefly discuss the application to two-field Horndeski and DHOST models. In Section \ref{sec:Single_scalar_field_Higher_Derivatives} we revisit the single scalar disformal transformations with higher derivatives and generalize the conditions derived in Ref.~\cite{auth:Takahashi_Matter_Coupling}. We also examine other ways to achieve consistent higher derivative modified gravity theories by using ansätze as \cite{Takahashi:2023vva} and \cite{Domenech:2023ryc}. We then extend the discussion to single vector fields with higher derivatives in section \ref{sec:single_vector} demonstrating that disformal transformation including the field strength \cite{DeFelice:2019hxb,Gumrukcuoglu:2019ebp} are highly restricted and propose a new class of transformations for higher derivative Proca theories without ghost degrees of freedom. Our results are summarized in section \ref{sec:Conclusion}, and some calculations are included in the appendix.  Throughout the paper, we use natural units $\hbar = c = M_{\rm pl}=1$ and use the mostly positive metric signature $(-,+,+,+)$.

\section{Preliminaries: transformation of the Dirac action}
\label{sec:Preliminaries}

Our starting point is a general (healthy) modified gravity theory, which depends on the metric and various fields, without the presence of Ostrogradsky ghosts. Standard (or other) matter fields are then coupled to a disformally related metric, which we refer to as disformal coupling. The disformal transformation relating both metrics depends in general on an arbitrary number of scalar, vector, tensor or even spinor fields. However, for simplicity, we will focus on the disformal transformations depending on a number ``I'' of scalar fields $\phi^{I}$ and on a number ``J'' of vector fields $\mathbf{B}^{J}$, and their derivatives, in what follows. We define disformal coupling as consistent, if it does not introduce new Ostrogradsky ghosts and leaves the original number of degrees of freedom unchanged. Such consistency condition is equivalent as checking whether the new modified gravity theory after performing the inverse disformal transformation does not have ghost degrees of freedom if standard matter is minimally coupled.

The action under consideration can be schematically expressed as
\begin{align}
    S = S_{\rm grav}[g_{\mu\nu}, \phi^{I}, \mathbf{B}^{J}] + S_m[ \tilde g_{\mu\nu}, \Psi^{M}, \mathbf{A}^{N}, \varphi]\,,
\end{align}
where $\Psi^{M}$ denotes ``M'' number of spinors, $\mathbf{A}^{N}$ are ``N'' number of gauge vector fields and $\varphi$ the Higgs field of the standard model. In the simplest case, we take the gravity sector as the Einstein-Hilbert action plus canonical kinetic terms for the disformal fields, that is $\phi^{I}$ and $\mathbf{B}^{J}$, respecting the same symmetries as the disformal couplings. However, the analysis can be straightforwardly expanded to non-minimal couplings of the disformal fields to gravity, such as in Horndeski and generalized Proca theories, as long as the equation of motion in the gravity sector are of second order. 

To be more concrete, let us show an example of an inconsistent disformal coupling with a single vector field. An action given by
\begin{align}
    S = S_{\rm EH} + \sqrt{-g} (F_B)_{\mu\nu} (F_B)^{\mu\nu} + S_M[ (g^{\mu\nu} + B_{\mu} B_\nu),\Psi^{M}, \mathbf{A}^{N},\varphi]\,,
\end{align}
where $(F_B)_{\mu\nu} = 2 \nabla_{[\mu} B_{\nu]}$ is the field strength tensor for the disformal field $B_\mu$, 
would be inconsistent as the disformal coupling breaks the $U(1)$ symmetry of the gauge field $B_\mu$ and, therefore, leads to an additional degree of freedom after coupling the matter sector. The disformal coupling would be only consistent if the starting action for the disformal field already breaks the $U(1)$ gauge symmetry.

\subsection{Disformal Transformation}

Let us briefly review a general disformal transformation of the Dirac action, closely following Ref.~\cite{auth:us}, and discuss the implications for the coupling to the fermions. We denote the disformal transformations of the local tetrads in general as
\begin{align}\label{eq:tetrads}
    \tilde e^A_\mu =  C \left({\cal U}\right)^{\nu}_{\mu}  e^A_\nu = C \,( \delta_\mu^\nu + {L}_\mu^\nu )  \,e^A_\nu, \quad{\rm and}\quad
    \tilde e^\mu_A =  C^{-1} \left({\cal U}^{-1}\right)^{\mu}_{\nu} e^\mu_A = C^{-1} ( \delta^\mu_\nu + {P}^{\mu}_{\nu} ) \,e^\nu_A\,,
\end{align}
where $C$, $L^\mu_\nu$ and $P^\mu_\nu$ are arbitrary spacetime functions. We assume that the tensor $\mathbf{L}$ is specified and, therefore, the inverse matrix $\mathbf{{\cal U}^{-1}}$, and equivalently $\mathbf{P}$, can be obtained by using the Cayley-Hamilton theorem. Namely, in matrix notation, one has that
 \begin{equation}
            \label{eq:Expression_P}
                \mathbf{P} =  \frac{1}{\det\mathcal{U}}\left[ -  \mathbf{L}^3 + ( 1 + \Tr \mathbf{L})  \mathbf{L}^2 + b_1 \mathbf{L} + b_0 \, \mathds{1} \right] \,,
            \end{equation}
        where
        \begin{align}
            b_1 =&  - \frac{1}{2} \left( 2 + \Trace \mathbf{L} (2 + \Trace \mathbf{L}) - \Trace \mathbf{L}^2 \right)  ~, \\
            b_0 =& \frac{1}{24} \left( - ( \Tr \mathbf{L})^4  - 8 \Tr \mathbf{L} \Tr \mathbf{L}^3 + 6 \Tr \mathbf{L}^4 + 6 (\Tr \mathbf{L})^2 \Tr \mathbf{L}^2 - 3  (\Tr \mathbf{L}^2 )^2 \right) ~, \\
        \end{align}
        and 
            \begin{align}
                \det \mathcal{U} =& 1 + \Tr \mathbf{L}  \left( 1 + \frac{1}{2} \Tr \mathbf{L} + \frac{1}{6} ( \Tr \mathbf{L})^2 + \frac{1}{24} ( \Tr \mathbf{L})^3 \right) + \frac{1}{3} \Tr \mathbf{L}^3  ( 1 + \Tr \mathbf{L}) - \frac{1}{4} \Tr \mathbf{L}^4  \nonumber \\
                & + \frac{1}{8} \Tr \mathbf{L}^2 \left( -4 - 4 \Tr \mathbf{L} - 2 (\Tr \mathbf{L})^2 + \Tr \mathbf{L}^2 \right) \,.
              \end{align} 
Note that, as shown in Ref.~\cite{auth:us}, one could, without loss of generality, assume that $\mathbf{L}$ is symmetric and traceless as any antisymmetric and trace component can be removed via local Lorentz transformations of the tetrads. However, for convenience, we do not assume any symmetry for $\mathbf{L}$. We will specify its form in terms of given fields later.

For easier comparison with the literature in disformal transformations, we write down the corresponding disformal transformation in terms of the spacetime metrics, which is given by
\begin{align}\label{eq:metricresulting}
    \tilde g_{\mu\nu} = \eta_{AB}  \tensor{\tilde e}{^A_\mu}  \tensor{\tilde e}{^B_\nu} = C^2 \left( g_{\mu\nu} + 2 L_{(\mu\nu)} +g_{\alpha\beta} {L}^\alpha_\mu {L}^\beta_\nu  \right) \,,
\end{align}
where indices in parenthesis denote normalized symmetrization, i.e. $L_{(\mu\nu)}=\tfrac{1}{2}(L_{\mu\nu}+L_{\nu\mu})$, where indices are lowered and raised with the metric $g_{\mu\nu}$, namely $L_{\mu\nu}=g_{\mu\alpha}L^\alpha_\nu$. From Eq.~\eqref{eq:metricresulting} we identify $C$ as the conformal factor and the combination involving $L^\mu_\nu$ is related to the pure disformal transformation.

We now turn to the transformation of the Dirac action. We consider a massive Dirac fermion in the hermitian formulation, minimally coupled to the tilded tetrads \eqref{eq:tetrads}, whose action reads
\begin{align}
\label{eq:Dirac_Lagrangian}
    S_{\rm Dirac} = - \int \md^4x\, \det \tilde e\, \Big[ \frac{1}{2} (\tilde{\overline \Psi} \gamma^\mu \tilde{\overset{\leftrightarrow}{ \nabla}}_\mu \tilde \Psi) +  \tilde{\overline \Psi}   m \tilde \Psi \Big]\,,
\end{align}
where $\overline \Psi=\Psi^\dag i \gamma^0$ is the Dirac adjoint spinor, the tilde denotes that they are minimally coupled to $\tilde g$ (also on top of the fermions),  and $\tilde{\overset{\leftrightarrow}{\nabla}}_\mu$ is defined through
\begin{align}
     (\tilde{\overline \Psi} \gamma^\mu \tilde{\overset{\leftrightarrow}{\nabla}}_\mu \tilde \Psi) = \tilde{\overline \Psi} \gamma^\mu \tilde \nabla_\mu \tilde \Psi - \tilde \nabla_\mu(\tilde{\overline \Psi} ) \gamma^\mu \tilde\Psi\,.
\end{align}
The gamma matrices with greek indices should be understood as $\gamma^\mu=e^\mu_A\gamma^A$, where $\gamma^A$ are the standard Dirac matrices.
Lastly, the covariant derivative of the spinor field (now without tildes to reduce notation density) is given by
\begin{align}
            \nabla_\mu \Psi = \partial_\mu \Psi + \frac{i}{2} \left({\omega}_{AB}\right)_\mu \Sigma^{AB} \Psi\quad {\rm and}\quad \nabla_\mu \overline\Psi = \partial_\mu \overline \Psi - \frac{i}{2} \left({\omega}_{AB}\right)_\mu \Sigma^{AB} \overline\Psi\,,
\end{align}
where $\Sigma^{AB}$ is the commutator of the gamma matrices, given by
\begin{align}
     \Sigma^{AB} = -\frac{i}{4} [ \gamma^A, \gamma^B ]\,,
\end{align}
and $\left({\omega}_{AB}\right)_\mu$ are the components of spin connection, defined by
 \begin{align}
        \left({\omega}_{AB}\right)_\mu = \tensor{e}{_C^\mu} \tensor{e}{_[_A^\nu} \left( \partial_\mu \tensor{e}{_B_]_\nu} - \tensor{e}{_B_]_\lambda} \Gamma^\lambda_{\mu\nu} \right)\equiv \tensor{e}{_C^\mu} \tensor{e}{_[_A^\nu}\nabla_\mu\tensor{e}{_B_]_\nu}\,.
\end{align}

The disformal transformation of the spin connection and the Dirac Lagrangian is derived in detail Ref.~\cite{auth:us}. For the readers convenience, we also provide it in App.~\ref{sec:Transformation_Tensor_Components}. Here, we simply write down the final result, which, after the transformation, reads\footnote{Note that in Ref.~\cite{auth:us} we worked in "mostly minus" convention. Here we used "mostly plus" instead. One is related to each other by $\gamma^\mu_{\cite{auth:us}}=i \gamma^\mu_{\rm here}$. We kept the same definition of $\gamma^5$, namely $\gamma^5=i\gamma^0\gamma^0\gamma^1\gamma^2\gamma^3$} 
\begin{align}
    \label{eq:Transformed_Dirac_Lagrangian}
    S_{\rm Dirac} = - \int \md^4x\, \det e\, \Big[ \frac{1}{2} \tensor{({\cal U}^{-1})}{^\mu_\nu} (\bar \Psi \gamma^\nu \overset{\leftrightarrow}{\nabla}_\mu \Psi) +  \bar \Psi C  m \Psi + i  M_\mu J^\mu_5    \Big] \,,
\end{align}
where $J_5^\alpha = \bar \Psi  \gamma^\alpha \gamma_5 \Psi$ is the axial current and we have rescaled the spinor field as
\begin{align}
    \label{eq:Spinor_rescaling}
    \Psi= C^{3/2} \sqrt{\det ({\cal U} )} \,\tilde\Psi\,.
\end{align}
We also defined for compactness
\begin{align}
    M^\alpha = & \frac{1}{4} \tensor{\epsilon}{^\alpha^\beta^\gamma^\delta} \tensor{(\mathcal{U}^{-1})}{^\mu_\delta}  \tensor{(\mathcal{U}^{-1})}{^\nu_\gamma} \nabla_\mu \tensor{\mathcal{U}}{_\beta_\nu}  \,.
     \label{eq:Definition_M_A}  
\end{align}
From Eq.~\eqref{eq:Transformed_Dirac_Lagrangian} we see that, while a conformal transformation only rescales the fermion mass, a disformal transformation modifies the kinetic term and introduces a coupling to the axial current via $M_\mu$. Such coupling to the axial current can be traced back to the possible presence of spacetime vorticity in the tilde metric \eqref{eq:metricresulting}. From Eq.~\eqref{eq:Definition_M_A}, we also see that $M_\mu$ depends on derivatives of ${\cal U}$. It is precisely this coupling which can prevent a consistent matter coupling and revive the Ostrogradsky ghost as discussed in \cite{auth:Takahashi_Matter_Coupling}. 

For later purposes it is important to note that the coupling to the axial current $J_5$ has a particular structure. Namely, after integration by parts, the terms with the covariant derivative acting on $J_5$ vanish identically due to the internal symmetries of $M_\mu$.\footnote{One can check that integrating by parts is equivalent to using the identity
\begin{align}
    \mathcal{U}^{-1} \nabla \mathcal{U} = - \mathcal{U} \nabla {\cal U}^{-1}\,.
\end{align}
} In short, $M_\mu$ never includes total derivative terms. They vanish after contraction.

\section{Two scalar fields}
\label{sec:Two_scalar_fields}
To get a first understanding of when the new coupling to the axial current can prevent a consistent matter coupling, we discuss a simple example of a disformal transformation depending on two scalar fields $\phi^{I}$ with $\phi^{I} \in \{\phi, \chi\}$,  
\begin{align}\label{eq:cIJs}
    L_{\mu\nu} = \sum_{I,J=1}^2 c_{IJ} \phi^{I}_\mu \phi^{J}_\nu
\end{align}
where we defined $\phi^{I}_\mu = \nabla_\mu \phi^{I}$ and $c_{IJ}$ are generic functions of the two scalar fields and their derivatives, $c_{IJ} = c_{IJ}(\phi,\chi, X,Q,Z)$ with $X = \phi_\mu \phi^\mu$, $Q= \phi_\mu \chi^\mu$ and $Z = \chi_\mu \chi^\mu$.  For simplicity, we assume that $c_{IJ}$ is symmetric.

For the case at hand, the inverse matrix $\mathbf{{\cal U}^{-1}}$ or equivalently $\mathbf{P}$ is then given by
\begin{align}\label{eq:PfordIJ}
    P_{\mu\nu} = \sum_{I,J} d_{IJ} \phi^{I}_\mu \phi^{J}_\nu\,,
\end{align}
where the coefficients $d_{IJ}$ are found using Eq.~\eqref{eq:Expression_P} and explicitly given by
\begin{align}
    d_{11} &= - \frac{c_{11} (1 + c_{22} Z) - c_{12}^2 Z }{T} \,,\\
    d_{12} &= - \frac{c_{12} (1 + c_{12} Q) - c_{11} c_{22} Q }{T}\,, \\
    d_{22} &= - \frac{c_{22} (1 + c_{11} X) - c_{12}^2 X }{T}\,, 
\end{align}
where we defined
\begin{align}
    T =& 1 - ( c_{11} X + c_{22} Z + 2 c_{12} Q) + ( X Z - Q^2) (c_{11} c_{22} - c_{12}^2 )\,.
\end{align}
We later present the transformation for vector fields. 

For later purposes, the disformal metric transformation can be expressed as
\begin{align}\label{eq:metricafterGIJ}
    \tilde g_{\mu\nu} = C^2 \left( g_{\mu\nu} + {\cal G}_{IJ} \phi^{I}_\mu \phi^{I}_\nu  \right)
\end{align}
where the field-space metric ${\cal G}_{IJ}$ is given by 
\begin{align}
    \label{eq:G_IJ_in_terms_c_IJ}
   {\cal G}_{IJ} = 2c_{IJ} + c_{IK} c_{JL} {\cal X}^{KL}
\end{align}
where we introduced  ${\cal X}^{IJ} = \phi^{I}_\mu \phi^{{J}\mu}$. 
Such disformal transformations has been studied in details in the context of multi-field DBI models. It is interesting to note that in order to have ${\cal G}_{IJ}={\cal G}_{IJ}(\phi,\chi)$, which only depends on the scalar fields and not their derivatives, $c_{IJ}$ has to depend on derivatives, $c_{IJ}=c_{IJ}(\phi,\chi,X,Q,Z)$. In what follows, we first study the axial coupling in detail, derive the necessary degeneracy conditions and lastly explore a subset of two-field Hordenski theories.

\subsection{Axial coupling}
    
Let us expand now the coupling to the axial current \eqref{eq:Definition_M_A}. Using that  $\nabla_{[\mu} \nabla_{\nu]}\phi^I =0$, the expression for $M_\alpha$ simplifies to
    \begin{align}\label{eq:Malpha1}
        M_\alpha = & \frac{1}{4} \tensor{\epsilon}{_\alpha^\beta^\gamma^\delta} \tensor{P}{^\mu_\gamma} \Big[ \tensor{P}{^\nu_\delta} c_{IJ} \phi^J_\mu   \nabla_\nu  \phi^I_\beta + \nabla_\delta (c_{IJ} \phi^J_\mu ) \phi^I_\beta  \Big] \nonumber \\
        =&  \frac{1}{4} \tensor{\epsilon}{_\alpha^\beta^\gamma^\delta} d_{KL} \phi^L_\gamma \Big[   c_{IJ} d_{MN} {\cal X}^{KJ}   \phi^{M\nu} \phi^N_\delta \nabla_\nu \phi^I_\beta   +   \phi^{K\mu}  \nabla_\delta(c_{IJ} \phi^J_\mu ) \phi^I_\beta \Big] \nonumber \\
        =&  \frac{1}{4} \tensor{\epsilon}{_\alpha^\beta^\gamma^\delta}   d_{KL} \phi^L_\gamma \phi^I_\beta \Big[ -   c_{NJ} d_{MI} {\cal X}^{KJ}    \phi^{M\nu} \nabla_\delta \phi^N_\nu   +   c_{IJ}  \phi^{K\mu} \nabla_\delta \phi^J_\mu   +  {\cal X}^{KJ}    \nabla_\delta c_{IJ}  \Big]\,,
    \end{align}
    where in the first step we used the explicit expression for $\tensor{P}{^\nu_\mu}$, Eq.~\eqref{eq:PfordIJ}, and in the second step we use the symmetries of the levi-civita tensor to pull out a common $\phi^I_\beta$ factor.
    See how $M_\alpha$ depends on the second derivatives of the scalar field leading in general to higher order equation of motions. 

    After expansion in field components, the expression \eqref{eq:Malpha1} for $M_\alpha$ , reads
    \begin{align}\label{eq:Malpha2}
        M_\alpha = & \frac{1}{4} \tensor{\epsilon}{_\alpha^\beta^\gamma^\delta}  \phi_\beta \chi_\gamma \Big[- b_1  \frac{1}{2} \nabla_\delta X - b_2  \chi^\nu  \nabla_\delta \phi_\nu - b_3  \phi^\nu  \nabla_\delta \chi_\nu - b_4 \frac{1}{2} \nabla_\delta Z \nonumber \\
        & + \nabla_\delta c_{11} (d_{12} X + d_{22} Q ) + \nabla_\delta c_{12} (- d_{11} X  + d_{22 } Z) - \nabla_\delta c_{22} (d_{11} Q + d_{21} Z) \Big]\,,
    \end{align}
    where we defined
    \begin{align}\label{eq:bs}
        b_1 = & \Big( (d_{12} c_{11} X + (d_{12} c_{12} + d_{22} c_{11} )Q + d_{22} c_{12} Z ) d_{11}  \nonumber \\
        & - (d_{11} c_{11} X + (d_{11} c_{12} + d_{21} c_{11} )Q + d_{12} c_{12} Z ) d_{12} - ( c_{11}  d_{12} - c_{21} d_{11} ) \Big)\,, \\
        b_2 =& \Big( (d_{12} c_{11} X + (d_{12} c_{12} + d_{22} c_{11} )Q + d_{22} c_{12} Z ) d_{21}  \nonumber \\
        & - (d_{11} c_{11} X + (d_{11} c_{12} + d_{21} c_{11} )Q + d_{12} c_{12} Z ) d_{22} - ( c_{11}  d_{22} - c_{21} d_{21} ) \Big) \,,\\
        b_3 = & \Big( (d_{12} c_{21} X + (d_{12} c_{22} + d_{22} c_{21} )Q + d_{22} c_{22} Z ) d_{11}  \nonumber \\
        & - (d_{11} c_{21} X + (d_{11} c_{22} + d_{21} c_{21} )Q + d_{12} c_{22} Z ) d_{12} - ( c_{12}  d_{12} - c_{22} d_{11} ) \Big) \,,\\
        b_4= & \Big( (d_{12} c_{21} X + (d_{12} c_{22} + d_{22} c_{21} )Q + d_{22} c_{22} Z ) d_{21}  \nonumber \\
        & - (d_{11} c_{21} X + (d_{11} c_{22} + d_{21} c_{21} )Q + d_{12} c_{22} Z ) d_{22} - ( c_{12}  d_{22} - c_{22} d_{21} ) \Big)\,.
    \end{align} 
    Since we are interested in the possible appearence of higher derivatives, we now neglect the dependence in $\phi$ and $\chi$, and consider that the $c_{IJ}$ are generic functions of $X$, $Q$ and $Z$, that is $c_{IJ}=c_{IJ}(X,Q,Z)$. Then, Eq.~\eqref{eq:Malpha2} becomes
    \begin{align}\label{eq:Malpha3}
        M_\alpha = \frac{1}{4} \tensor{\epsilon}{_\alpha^\beta^\gamma^\delta}  \phi_\beta \chi_\gamma  \Big[ h_1 \frac{1}{2} \nabla_\delta X + h_2 \chi^\nu \nabla_\delta \phi_\nu + h_3 \phi^\nu \nabla_\delta \chi_\nu + h_4 \frac{1}{2} \nabla_\delta Z \Big]\,,
    \end{align}
    where the new coeffcieints depend only on $X$, $Q$, $Z$, and are explicitly given by
    \begin{align}\label{eq:hs}
        h_1 = & - b_1 + 2 c_{11,X} (d_{12} X + d_{22} Q) + 2 c_{12,X} (-d_{11} X + d_{22} Z) - 2  c_{22,X} (d_{11} Q + d_{12} Z), \\
        h_2 = &  - b_2 + c_{11,Q} (d_{12} X + d_{22} Q) + c_{12,Q} (-d_{11} X + d_{22} Z) - c_{22,Q} (d_{11} Q + d_{12} Z), \\
        h_3 =&  - b_3 + c_{11,Q} (d_{12} X + d_{22} Q) + c_{12,Q} (-d_{11} X + d_{22} Z) - c_{22,Q} (d_{11} Q + d_{12} Z), \\
        h_4 =&  - b_4 +2 c_{11,Z} (d_{12} X + d_{22} Q) + 2 c_{12,Z} (-d_{11} X + d_{22} Z) - 2 c_{22,Z} (d_{11} Q + d_{12} Z)\label{eq:h4}\,.
    \end{align}
    From Eq.~\eqref{eq:Malpha3}, we see that $M_\alpha$ contains in general second time derivatives of the scalar fields. In general, the kinetic matrix is non-degenerate and the equations of motion will contain third time derivatives leading to Ostrogradsky ghosts. Note that the terms proportional to $h_2$ and $h_3$ in Eq.~\eqref{eq:Malpha3} cannot be written, in general, as a single combination of derivative of $Q$. We will come to this shortly.

    To demonstrate the presence of third time derivatives, let us consider one simple example where we fix $c_{22}= c_{12}^2/c_{11}$ and $c_{12}=\mathrm{const.}$ but $c_{11}=c_{11}(X)$. 
    For such choice, we obtain $b_1=b_2=b_3=b_4=0$, which helps to drastically simplify the equations. Then, one finds that equations of motion of the scalar field $\phi$ will contain terms proportional to
    \begin{align}\label{eq:dsdphi1}
        \frac{\delta S}{\delta \phi} \supset    \frac{1}{8} J_5^\alpha \tensor{\epsilon}{_\alpha^\beta^\gamma^\delta} \phi_\beta \chi_\gamma \Big[ ( 2 h_{1,X} \phi^\sigma + h_{1,Q} \chi^\sigma ) \nabla_\sigma  \nabla_\delta X - 2 \phi^\sigma \nabla_\sigma \nabla_\delta h_1  \Big]\,,
    \end{align}
    where by ${\delta S}/{\delta \phi}$ we mean the full variation, i.e. also variation with respect to derivatives of the field. In Eq.~\eqref{eq:dsdphi1} we only selected the terms which contain third time derivatives of $\phi$, here through the term $\nabla_\sigma  \nabla_\delta X$. Note, however, that in the case where one of the scalar fields (or a linear combination of them) is time-like and we identified it as the time direction, e.g. $\chi_\mu \sim n_\mu$ with $n_\mu n^\mu=-1$, the equations of motion only depend on second time derivatives of $\phi$, due to the antisymmetric structure of the kinetic matrix. This implies that the disformal coupling with two scalar fields generically belongs to the U-DHOST theories \cite{DeFelice:2018ewo,DeFelice:2021hps}. This will not be the case as soon as we consider three or more scalar fields as then there are more independent spatial directions. In general, though, if we do not choose such special slicing (commonly referred to as the the unitary gauge), we need to impose degeneracy conditions to avoid the presence of Ostrogradsky ghosts. This later approach is what we will pursue below.

    \subsection{Degeneracy conditions}
    To avoid the presence of Ostrogradsky ghosts in Eq.~\eqref{eq:Transformed_Dirac_Lagrangian}, we need to impose degeneracy conditions. Under general assumptions, the two scalar fields are independent from each other. Although there might be solutions for which this is not the case, such as when both fields are functions of time only, we will not be concerned in these particular situations. 

    We start by exploiting the absence of a term $\nabla_\delta Q$ in Eq.~\eqref{eq:Malpha3}. Focusing on the terms with $h_2$ and $h_3$, we see that the equations of motion for $\phi$ and $\chi$ contain terms like
    \begin{align}
        \frac{\delta S}{\delta \phi} \supset & \frac{1}{8} J_5^\alpha \tensor{\epsilon}{_\alpha^\beta^\gamma^\delta} \phi_\beta \chi_\gamma  (h_3 - h_2) \phi \nabla_\sigma  \nabla^\sigma \nabla_\delta \chi , \\
        \frac{\delta S}{\delta \chi} \supset &  \frac{1}{8} J_5^\alpha \tensor{\epsilon}{_\alpha^\beta^\gamma^\delta} \phi_\beta \chi_\gamma  (h_2 - h_3) \chi \nabla_\sigma  \nabla^\sigma \nabla_\delta \phi\,,
    \end{align}
    which have a different structure than those in e.g. Eq.~\eqref{eq:dsdphi1}. Namely, there is a Dalambertian, that is $\Box\equiv\nabla_\sigma  \nabla^\sigma$, acting directly on the fields. We will come back to the other terms with third derivatives in a moment. But, in a nutshell, one can check that in general there is are cancellations between terms proportional to $\chi \nabla_\sigma  \nabla^\sigma \nabla_\delta \phi$ and second derivatives of $X$, $Q$ or $Z$, e.g. $ \chi^\sigma \nabla_\sigma  \nabla_\delta X$. Therefore, the first degeneracy condition is given by 
    \begin{align}\label{eq:degeneracy1}
    h_2=h_3\,,
    \end{align}
    which is equivalent to $b_2 = b_3$. In other words, the action \eqref{eq:Transformed_Dirac_Lagrangian} should first depend only on derivatives of $X$, $Q$ and $Z$.

    Before presenting the requirements on the disformal coefficients, let us impose the condition $h_2 = h_3$ back into $M_\alpha$ \eqref{eq:Malpha3}. Doing so, we obtain
    \begin{align}\label{eq:Malpha4}
        M_\alpha = \frac{1}{4} \tensor{\epsilon}{_\alpha^\beta^\gamma^\delta}  \phi_\beta \chi_\gamma  \Big[ h_1 \frac{1}{2} \nabla_\delta X + h_2 \nabla_\delta Q + h_4 \frac{1}{2} \nabla_\delta Z \Big]\,.
    \end{align}
    As $M_\alpha$ depends on derivatives of $X$, $Q$ and $Z$, it may be possible to introduce a total differential given by 
     \begin{align}\label{eq:H_differential}
        \md H = \frac{1}{2} h_1 \md X + h_2 \md Q + \frac{1}{2} h_4 \md Z\,,
    \end{align}
    which would be integrable only under certain conditions on $h_1$, $h_2$ and $h_4$,  that is if
    \begin{align}\label{eq:integrabilitycond}
    h_{1,Q}=2h_{2,X}\quad,\quad h_{4,Q}=2h_{2,Z}\quad {\rm and}\quad h_{1,Z}=h_{4,X}\,.
    \end{align}
    Note that we could also include arbitrary terms proportional $d_1 \md \phi$ and $d_2 \md \chi$ in Eq.~\eqref{eq:H_differential}, as they vanish after contractions with the Levi-Civita tensor. Let us show that the integrability conditions for \eqref{eq:H_differential} are precisely those that also lead to the absence of higher time derivatives.

    We proceed as before and assume that $\phi$ and $\chi$ are in general independent. We then focus on their equations of motion. All the terms with third order derivatives are given by
    \begin{align}
        \frac{\delta S}{\delta \phi} \supset &   \frac{1}{8} J_5^\alpha \tensor{\epsilon}{_\alpha^\beta^\gamma^\delta} \phi_\beta \chi_\gamma \Big[ (h_{1,Q} - 2  h_{2,X}) \chi^\sigma\nabla_\sigma \nabla_\delta X + (4 h_{2,X} \phi^\sigma - 2 h_{1,Q} \phi^\sigma ) \nabla_\sigma \nabla_\delta Q \nonumber \\
        & + (2 h_{4,X} \phi^\sigma - 2 \phi^\sigma h_{1,Z} - 2 \chi^\sigma h_{2,Z} + h_{4,Q} \chi^\sigma ) \nabla_\sigma \nabla_\delta  Z \Big]\,,\\
        \frac{\delta S}{\delta \chi} 
        \supset &  \frac{1}{8} J_5^\alpha \tensor{\epsilon}{_\alpha^\beta^\gamma^\delta} \phi_\beta \chi_\gamma \Big[   ( (2 h_{1,Z} -2 h_{4,X} ) \chi^\sigma + (h_{1,Q} - 2 h_{2,X} ) \phi^\sigma ) \nabla_\sigma \nabla_\delta X +   (4h_{2,Z} - 2 h_{4,Q}  ) \chi^\sigma     \nabla_\sigma \nabla_\delta Q \nonumber \\
        & +   (h_{4,Q} - 2 h_{2,Z} ) \phi^\sigma  \nabla_\sigma \nabla_\delta Z \Big]\,.
    \end{align}
    If the fields are independent, then each of the terms in brackets must vanish independently. These requirements precisely coincides with the conditions under which $dH$ \eqref{eq:H_differential} is integrable, Eq.~\eqref{eq:integrabilitycond}.

    After imposing all degeneracy conditions, we conclude that $M_\alpha$ \eqref{eq:Malpha4} must be
    \begin{align}\label{eq:MalphaandH}
        M_\alpha = \frac{1}{4} \tensor{\epsilon}{_\alpha^\beta^\gamma^\delta}  \phi_\beta \chi_\gamma \nabla_\delta H \,,
    \end{align}
    namely only a function of $\nabla_\delta H$. However, the form of Eq.~\eqref{eq:MalphaandH} is inconsistent with the original expression \eqref{eq:Definition_M_A} if $H$ is a function of $X$, $Q$ and $Z$. If it were, we could then write $ M_\alpha$ as a total derivative, that is
     \begin{align}\label{eq:MalphaandH2}
        M_\alpha = \frac{1}{4} \tensor{\epsilon}{_\alpha^\beta^\gamma^\delta}  \phi_\beta \chi_\gamma \nabla_\delta H= \frac{1}{4} \tensor{\epsilon}{_\alpha^\beta^\gamma^\delta} \nabla_\delta\left( \phi_\beta \chi_\gamma H \right)\,.
    \end{align} 
    But, as we discussed below Eq.~\eqref{eq:Definition_M_A}, this is not possible. Thus, the only consistent solution is that 
    \begin{align}\label{eq:degeneracy2}
    H=H(\phi,\chi)\Rightarrow h_1=h_2=h_4=0\,.
    \end{align}

Let us now study the consequences of the degeneracy conditions, Eqs.~\eqref{eq:degeneracy1} and \eqref{eq:degeneracy2}, for the disformal coefficients $c_{IJ}$, introduced in Eq.~\eqref{eq:cIJs}. First, using Eqs.~\eqref{eq:hs} and \eqref{eq:bs}, we see that the condition $h_2=h_3$ \eqref{eq:degeneracy1} has three possible solutions. The solutions, which we label as $(i)$, $(ii)$ and $(iii)$, are given by
    \begin{align}
        c^{(i)}_{22} = & \frac{c_{12}^2}{c_{11}}, \label{eq:sol1} \\
        c^{(ii)}_{22} = & - \frac{1}{Z}\left({1 + X c_{11} + 2Q c_{12} }\right), \label{eq:sol2} \\
        c^{(iii)}_{22} = & \frac{1}{c_{11}}\left({c_{12}^2 + \frac{1}{-Q^2 + X Z}}\right)\,. \label{eq:sol3}
    \end{align}
   As we shall see, class $(i)$ is effectively a disformal transformation along a single effective direction in spacetime. This is not the case for classes $(ii)$ and $(iii)$, which render without a smooth limit of vanishing disformal transformation.  Let us consider them separately below. 

    \subsubsection{Class \textit{(i)}}

    Using \eqref{eq:sol1} for $c^{(i)}_{22}$, together with \eqref{eq:degeneracy2}, leads to $b_1=b_2 =b_3=b_4=0$. In that case, the degeneracy conditions drastically simplify to three conditions which have a structure given by
    \begin{align}
    \label{eq:Constraint_Equation}
        (X c_{11}^2 + 2 Q c_{11} c_{12} + Z c_{12}^2) \left(-c_{12} \frac{\partial c_{11}}{\partial {\cal X}} + c_{11} \frac{\partial c_{12}}{\partial {\cal X}}\right) = 0\,,
    \end{align}
    where ${\cal X}=\{X,Q,Z\}$. The solution, therefore, has two branches. One in which the second parenthesis vanishes and one in which the first parenthesis is zero.  Let us call them branch $(i1)$ and $(i2)$ respectively.

    The first branch $(i1)$ is given by solving the partial differential equation in \eqref{eq:Constraint_Equation}. This leads to
    \begin{align}
         c^{(i1)}_{12} = f(\phi,\chi)  c_{11}(\phi,\chi,X,Q,Z), \qquad c^{(i1)}_{22} = f^2(\phi,\chi) c_{11}(\phi,\chi,X,Q,Z)\,,
    \end{align}
    with $c_{11}(\phi,\chi,X,Q,Z)$ still an arbitrary function of the arguments. To see the implications of such conditions, let us write the metric transformation \eqref{eq:metricafterGIJ} explicitly for branch $(i1)$. We find that $\tilde g_{\mu\nu}$, is given by
    \begin{align}\label{eq:tildegmunuandD}
        \tilde g_{\mu\nu } = & C^2 \Big( g_{\mu\nu} + D  \phi_\mu \phi_\nu + 2 f D  \phi_{(\mu} \chi_{\nu)} + f^2 D \chi_{\mu} \chi_\nu  \Big) \,,
    \end{align}
    where we defined
    \begin{align}
        D = c_{11} (2 + c_{11} (X + 2 f Q + f^2 Z)) \,.
    \end{align}

   From Eq.~\eqref{eq:tildegmunuandD}, it is now straightforward to see why the disformal transformation does not lead to any issues, as the transformation can be expressed as an effective single-scalar field disformal transformation. Namely, the disformal transformation stretches the metric in a single effective direction. This is clear if we introduce a new scalar field $\Phi$ via
   \begin{align}\label{eq:Phi}
       \md \Phi = (\md \phi + f(\phi,\chi)  \md \chi ) g(\phi,\chi)\,,
   \end{align}
   where $g(\phi,\chi)$ can be any arbitrary function. Note, however, that the $\md \Phi$ \eqref{eq:Phi} is in general not integrable, which reflects the fact that the disformal transformation still depends on two scalar fields. $\Phi$ \eqref{eq:Phi} is integrable only if $f$ factorizes, i.e. $f= f_1 (\phi) f_2(\chi)$, and $g=1/f_1(\phi)$ is appropriately fixed.
   
   Last, let us shortly comment about the second branch, $(i2)$, which is given by setting the prefactor in \eqref{eq:Constraint_Equation} to zero. This results in
   \begin{align}\label{eq:conditioni2}
        c^{(i2)}_{12} = \frac{- Q c_{11} \pm \sqrt{(Q^2- X Z) c_{11}^2} }{Z}\,.
    \end{align}
    Note that this solution is only valid as long as $Q^2- X Z \geq 0$, which constrains the scalar fields. In particular, if the derivatives of both scalar fields are time-like, this condition is never fulfilled, unless the derivative of the two scalar-fields point exactly in the same direction. Therefore, this is in general not satisfied. Nevertheless, assuming the condition \eqref{eq:conditioni2} is fulfilled, we find that the tilde metric reads
    \begin{align}\label{eq:othertildeclassi2}
        \tilde g_{\mu\nu} = C^2 \left( g_{\mu\nu} + 2 c_{11} \left[  \phi_\mu \phi_\nu +  2 F \phi_{(\mu} \chi_{\nu)} + F^2  \chi_\mu \chi_\nu  \right] \right)\,,
    \end{align}
    where
    \begin{align}
        F = \frac{-Q \pm {\rm sgn}(c_{11}) \sqrt{Q^2- X Z} }{Z}\,.
    \end{align}
    Formally, we can again note that it is effectively a transformation along one direction. However, we are not be able to find a closed solutions as $F$ depends on the derivatives of the fields. 

   \subsubsection{Class \textit{(ii)} and \textit{(iii)}}

   Lastly, we look at the implications of conditions \eqref{eq:sol2} and \eqref{eq:sol3}, respectively corresponding to classes $(ii)$ and $(iii)$. As will be clear later, we investigate first the main properties of the disformal field space metric ${\cal G}_{IJ}$, related to $c_{IJ}$ by Eq.~\eqref{eq:G_IJ_in_terms_c_IJ}, only imposing \eqref{eq:sol2} and \eqref{eq:sol3} but not Eq.~\eqref{eq:degeneracy2}. We find that, since $c_{22}$ in both \eqref{eq:sol2} and \eqref{eq:sol3} has a term which is not proportional to $c_{11}$ nor $c_{12}$, there is also a term in ${\cal G}_{IJ}$ with the same properties. This implies that  classes $(ii)$ and $(iii)$ do not have a smooth limit to a vanishing disformal field space metric, i.e. ${\cal G}_{IJ}\rightarrow 0$. Even the trivial solution $X=Q=Z=0$ seems problematic as the coefficients diverge. In other words, such a disformal coupling can never be turned off. Thus, not to contradict observations today the disformal coupling should be highly suppressed, that is ${\cal G}_{IJ} \ll 1$. For this reason, we consider that classes $(ii)$ and $(iii)$ are not as interesting as class $(i)$. In addition to above-mentioned issue, the second degeneracy conditions, \eqref{eq:degeneracy2}, become complicated partial differential equations for the coefficients $c_{IJ}$, see Eqs.~\eqref{eq:hs}--\eqref{eq:h4}. Unfortunately, we did not find general solutions to such equations. We leave a more detailed study of classes $(ii)$ and $(iii)$ for future work.

   \subsection{Two-field Beyond Horndeski/ DHOST}

    In the previous section, we have shown that in order to have a consistent matter coupling the disformal transformation with two scalar fields has to point along one effective direction. With a consistent coupling, we can explore the corresponding modified gravity theory by performing the inverse disformal transformation.

    To avoid using tildes on top of every quantity in the transformed frame, we use, only in this subsection, a tilde for Einstein frame and no tilde for Jordan frame quantities. For further simplicity, we focus only on the pure disformal part of the metric transformation, which in class $(i)$ (see Eqs.~\eqref{eq:tildegmunuandD} and \eqref{eq:othertildeclassi2}) is given by 
    \begin{align}
        \tilde g_{\mu\nu} = g_{\mu\nu} + D(\phi,\chi,X,Q,Z) G_{IJ}(\phi,\chi) \phi^I_\mu \phi^J_\nu, \qquad \tilde g^{\mu\nu} = g^{\mu\nu} - \frac{D}{1 + D G_{KL} \phi^K_\mu \phi^{L\mu}} G_{IJ} \phi^{I\mu} \phi^{J\nu}\,,
    \end{align}
    where $G_{IJ}$ is the degenerate field space metric
    \begin{align}\label{eq:gijdegenerate}
        G_{IJ} = \delta_I^1 \delta_J^1 + 2 f \delta_{(I}^1 \delta_{J)}^2 + f^2 \delta_I^2 \delta_J^2  \,.
    \end{align}
    We found that, for calculations, it is more practical to introduce the vector $V_\mu = \phi_\mu + f \chi_\mu$ with which the metric transformation  simply reads
    \begin{align}
        \tilde g_{\mu\nu} = g_{\mu\nu} + D V_\mu V_\nu\,.
    \end{align}
    We can later expand the resulting action in terms of $\phi$ and $\chi$.

    In the following, let us focus on the case $D=D(\phi,\chi)$, which in the single scalar field case leads to Horndeski theory. We will later discuss the implications for the general case. In general, one could start from the two-field Horndeski model \cite{Horndeski:2024hee} as it leads to second order equation of motions. But, for simplicity, let us assume that gravity in the Einstein-frame is just given by the Einstein-Hilbert action. As we shall see, we find new terms even in this simple case.
       
    For the disformal transformation of the Ricci scalar, we use the relations \cite{BenAchour:2016cay}
    \begin{align}
        \tilde R =  \tilde R_{\mu\nu} \tilde g^{\mu\nu} = & \tilde g^{\mu\nu} \left( R_{\mu\nu} + C^\sigma_{\mu\rho} C^\rho_{\mu\sigma} - C^\rho_{\mu\nu} C^\sigma_{\rho\sigma} \right) + \tilde \nabla_\rho \xi^\rho\,,
    \end{align}
    where 
    \begin{align}
    C^\rho_{\mu\nu}=\tilde\Gamma^\rho_{\mu\nu}-\Gamma^\rho_{\mu\nu}=\frac{1}{2}\tilde g^{\rho\lambda}\left(2\nabla_{(\mu}\tilde g_{\nu)\lambda}-\nabla_{\lambda}\tilde g_{\mu\nu}\right)\,,
    \end{align}
    and
    \begin{align}
        \xi^\rho = \tilde g^{\mu\nu} C^\rho_{\mu\nu} - \tilde g^{\rho\mu} C^\nu_{\mu\nu}\,.
    \end{align}
    As we start from the Einstein-Hilbert term we can discard the total derivative term. Lastly, we also use the following identity,
    \begin{align}
        R_{\mu\nu} V^\mu V^\nu = V^\mu \nabla_\mu \nabla_\nu V^\nu - V^\mu \nabla_\nu \nabla_\mu V^\nu\,.
    \end{align}

    After some algebra, the disformal transformation of the Einstein-Hilbert action leads to
    \begin{align}\label{eq:twofieldEH}
        S_{\rm EH} = & \frac{1}{2} \int \md^4x\, \sqrt{-g}  \Big[  \sqrt{1+ D  G_{IJ} \phi^{I\mu} \phi_\mu^J}\, R + \frac{D}{\sqrt{1+ D  G_{IJ} \phi^{I\mu} \phi_\mu^J}} G_{IJ} \left( \phi^I_{\mu\nu} \phi^{J\mu\nu} - \Box \phi^I \Box\phi^J \right) \nonumber \\
        & + \frac{ (f_\chi D_\phi - f_\phi D_\chi )  ( Q^2 - X Z)}{\sqrt{1+ D  G_{IJ} \phi^{I\mu} \phi_\mu^J}} +  M_{IJK} \left( \delta^\mu_\delta \delta^\nu_\sigma  -\delta^\mu_\sigma \delta ^\nu_\delta \right) \phi^I_\mu \phi^{J\delta} \phi^{K \sigma }_\nu   \Big]~,
    \end{align}
   where we defined
    \begin{align}
        M_{IJK} = - \frac{ ( \delta_K^1 + f \delta_K^2 )}{\sqrt{1+ D  G_{IJ} \phi^{I\mu} \phi_\mu^J}} \Big[ \delta_I^1 \delta_J^1 D_\phi + 2 \delta_{(I}^1 \delta_{J)}^2 ( f D_\phi + D_\chi + 2 D f_\phi ) + \delta_I^2 \delta_J^2 (f D_\chi + 2 D f_\chi ) \Big]~.
    \end{align}
    Note that, while the first three terms in Eq.~\eqref{eq:twofieldEH} are part of the two-field Galileon model \cite{Padilla:2012dx}, the last term with $M_{IJK}$ is not. Yet, the equations of motion are still of second order. Indeed, such a term has been proposed in Ref.~\cite{Ohashi:2015fma} as a generalization of the two-field Galileon model. It is also present in \cite{Horndeski:2024hee} and correspond to $L_{12}$, $L_{13}$ and $L_{14}$. Here we showed for the first time that this type of term naturally arises if one considers a two-field disformal transformation with a consistent matter coupling. It is worth noting that that the new term is even present if $f=0$ in Eq.~\eqref{eq:gijdegenerate} as long as $D=D(\phi,\chi)$.

    In future works it would be interesting to study the disformal transformation of the full two-field Horndeski Lagrangian derived in \cite{Horndeski:2024hee} to explicitly check whether the equations of motion are remain of second order after the disformal transformation. It would also be interesting to explore whether there is a closure relation for the functional form of the action, as is for the single scalar field Horndeski model. Although naively expected, it is a priori not guaranteed that the transformed Lagrangian still has second order equation of motion in the Jordan frame. There will be, however, no ghost degree of freedom as the coupling to matter was consistent in the Einstein-frame.

    Lastly, we can consider the most general the ansatz for the consistent disformal transformation, that is $C=C(\phi,\chi,X,Q,Z)$ and $D=D(\phi,\chi,X,Q,Z)$. In that case, one would obtain a two-field DHOST model, where the equation of motions will in general not be anymore of second order. However, as the matter coupling is still consistent, and the equation of motions are second order in the Einstein frame, there will be internal constraint which guarantees that there are no ghost degrees of freedom in the Jordan frame. By iteratively adding the new operators starting from two-field Horndeski \cite{Horndeski:2024hee} one can explore a subset of the two field DHOST model. Note, that this approach is only consistent as long as we keep the direction of the disformal transformation fix and do not consider different $V_\mu$ at each step. The form of $V_\mu$ itself is, however, arbitrary. As the main purpose of this paper is the matter coupling, we leave these interesting directions as future work.

\section{Single scalar field with higher derivatives}
\label{sec:Single_scalar_field_Higher_Derivatives}

As a next step, let us explore the case with a single scalar field $\phi$ but with higher order spatial derivatives which have been extensively studied in the recent literature, see Refs.~\cite{Takahashi:2022mew,auth:Takahashi_Invertible_Disformal,auth:Takahashi_Matter_Coupling,auth:Ikeda_Consistency_Couplings,Takahashi:2023vva,Takahashi:2023jro}. 
In the following, we assume that the scalar field is time-like and defines a preferred time-slicing, i.e. $\nabla_\mu \phi \propto n_\mu$ where $n_\mu$ is the normal direction to the hypersurface of constant time. The metric can then be decomposed as
\begin{align}
    g_{\mu\nu} = h_{\mu\nu} - n_\mu n_\nu\,,
\end{align}
where $h_{\mu\nu}$ is the intrinsic metric of the hypersurface. As long as the disformal transformation is invertible, it can be used to construct new consistent modified gravity theories of gravity leading to a generalized class of U-DHOST. We first explore the case of a disformal transformation with second derivatives, then turn to arbitrary higher spatial derivatives and lastly to the special case of disformal invariant combinations.

\subsection{Second derivatives\label{sec:secondderivativesscalar}}

Disformal transformation with second derivatives of the scalar field have been studied in detail in Ref.~\cite{auth:Takahashi_Invertible_Disformal}. Here we generalize their work further by relaxing one of the conditions for a consistent matter coupling. We consider a disformal transformation, in a notation similar to Ref.~\cite{auth:Takahashi_Invertible_Disformal}, of the form
\begin{align}\label{eq:gsecondderivatives}
    \tilde g_{\mu\nu} = f_0 g_{\mu\nu} + f_1 \phi_\mu \phi_\nu +  2 f_2 \phi_{(\mu} {\cal X}_{\nu)} + f_3 {\cal X}_\mu {\cal X}_\nu\,,
\end{align}
where ${\cal X}_\mu = h_\mu^\nu \nabla_\nu X$ and $f_i$ are the arbitrary functions of $\phi$, $X$ and  ${\cal Z}= {\cal X}_\mu {\cal X}^\mu$, that is $f_i=f_i(\phi,X,{\cal Z})$. The conditions for invertibility of the metric are given in Ref.~\cite{auth:Takahashi_Invertible_Disformal}. Later, Ref.~\cite{auth:Takahashi_Matter_Coupling} argued that the coupling to fermions revives the Ostrogradsky ghost if $f_3 \neq 0$. However, as we we will show, although $f_3=0$ is a sufficient condition it is not necessary. Below, we derive the general conditions for a consistent matter coupling. 

For the transformation of the tetrad we compare Eqs.~\eqref{eq:metricresulting} and \eqref{eq:gsecondderivatives}, and identify that
\begin{align}\label{eq:Lmunusecondder}
    \tensor{L}{^\mu_\nu} = c_{00} \phi^\mu \phi_\nu +  c_{01} \phi^\mu {\cal X}_{\nu} + c_{10} {\cal X}^\mu \phi_\nu + c_{11} {\cal X}^\mu {\cal X}_\nu\,,
\end{align}
where $c_{00}$, $c_{01}$, $c_{10}$ and $c_{11}$ are arbitrary functions of $\phi$, $X$ and ${\cal Z}$. For easier comparison with the literature, the free functions $f_i$ in \eqref{eq:gsecondderivatives} are related to $c_{ij}$ via
\begin{align}
     f_0 = C^2 \,\phantom{\left( c_{01} + c_{10} + c_{00} c_{01} X + c_{01} c_{11} {\cal Z} \right)}\quad&,\quad   f_1 = C^2 \left( 2 c_{00} + c_{00}^2 X + c_{10}^2 {\cal Z}  \right)\,, \nonumber \\
    f_2 = C^2 \left( c_{01} + c_{10} + c_{00} c_{10} X + c_{01} c_{11} {\cal Z} \right)\quad&,\quad 
     f_3 = C^2 \left( 2 c_{11} + c_{11}^2 {\cal Z} + c_{01}^2 X  \right)\,.
    \label{eq:Definition_fi_metric}
\end{align}
As noted in Ref.~\cite{auth:Takahashi_Matter_Coupling}, one can use a Lorentz transformation to simplify the ansatz \eqref{eq:Lmunusecondder}. In general, one can set $c_{01}=c_{10}$ or $c_{01}=0$ without loss of generality. However, note that by fixing $c_{01}=0$ the transformation of the tetrad \eqref{eq:Lmunusecondder} acquires antisymmetric components. We found that antisymmetric components may be absorbed by a redefinition of the spinor. If this is not taken into account, the fixing of $c_{01}=0$ may lead to misinterpretations of the results. For this reason, we maintain a general ansatz and discuss Lorentz transformations later.

To compute the transformation of the Dirac action, we must also compute $\tensor{P}{^\mu_\nu}$ \eqref{eq:Expression_P}. Using $\tensor{L}{^\mu_\nu}$ as given in \eqref{eq:Lmunusecondder}, we find that
\begin{align}
    \tensor{P}{^\mu_\nu} = d_{00} \phi^\mu \phi_\nu +  d_{01} \phi^\mu {\cal X}_{\nu} + d_{10} {\cal X}^\mu \phi_\nu + d_{11} {\cal X}^\mu {\cal X}_\nu\,,
\end{align}
where  
 \begin{align}
        d_{00} = & - \frac{c_{00} + c_{00} c_{11} {\cal Z} - c_{01} c_{10} {\cal Z} }{T}, \\
        d_{01} =& - \frac{c_{01}}{T}, \\
        d_{10}=& - \frac{c_{10}}{T}, \\
        d_{11} =& - \frac{c_{11} + c_{00} c_{11} X - c_{01} c_{10} X }{T}, \\
        T= & 1+ c_{00} X + c_{11} {\cal Z} + X {\cal Z} (c_{00} c_{11} - c_{01} c_{10})\,.
\end{align}
As we did in previous sections, we require that the resulting action does not depend on second time derivatives of $\phi$, to avoid the revival of the Ostrogradsky ghost. We can then focus only on the terms with time derivatives of ${\cal X}_\mu$ in the axial coupling \eqref{eq:Definition_M_A}. Doing so, we obtain that 
\begin{align}\label{eq:Malphasecondderiv}
        M_\alpha \supset & - \frac{1}{4 X} \tensor{\epsilon}{_\alpha^\beta^\gamma^\delta}   \left( c_{11} {\cal Z} \left(- d_{11}   + d_{01} d_{10} X  - d_{00}  d_{11} X  \right) - d_{01} c_{10} X  \right) \phi_\delta {\cal X}_\gamma   \phi^\mu \nabla_\mu  {\cal X}_\beta  \nonumber \\
        = & - \frac{c_{01} c_{10} X + c_{11}^2 {\cal Z}}{4 X T} \tensor{\epsilon}{_\alpha^\beta^\gamma^\delta}  \phi_\delta {\cal X}_\gamma   \phi^\mu \nabla_\mu  {\cal X}_\beta\,.
    \end{align}
We proceed to discuss the conditions on $f_3$ for a consistent matter coupling. We first use a Lorentz transformation to fix $c_{01}=c_{10}$ and later we repeat the procedure fixing $c_{01}=0$ instead.

    \subsubsection{Symmetric ansatz}
    Without loss of generality let us fix $c_{01}=c_{10}$. In this case, the terms with time derivatives simplify to 
    \begin{align}
        M_\alpha \supset &  - \frac{c_{01}^2 X + c_{11}^2 {\cal Z}}{4 X T} \tensor{\epsilon}{_\alpha^\beta^\gamma^\delta} \phi_\delta {\cal X}_\gamma   \phi^\mu \nabla_\mu  {\cal X}_\beta \,.
        \label{eq:M_mu_symmetric_ansatz}
    \end{align}
    Therefore, in order to remove the time derivative of ${\cal X}_\mu$ we can impose the degeneracy condition
    \begin{align}\label{eq:c01sol}
        c_{01}^2 = - \frac{c_{11}^2 {\cal Z}}{X}\,.
    \end{align}
    Note that $c_{01}$ is real valued, since  ${{\cal X}_\mu}$ is spacelike while $\phi_\mu$ is time-like and, therefore, ${\cal Z}/X < 0$.
    By inserting \eqref{eq:c01sol} into Eq.~\eqref{eq:Definition_fi_metric}, we find that
    \begin{align}\label{eq:f3now}
    f_3 = 2 C^2 c_{11}\,.
    \end{align}
    Thus, there are disformal couplings to matter with $f_3\neq 0$ which are consistent. We also checked that one can construct invertible disformal transformations consistent with the degeneracy conditions and $f_3 \neq 0$ following \cite{auth:Takahashi_Invertible_Disformal}. Let us discuss below the differences of our work with Ref.~\cite{auth:Takahashi_Matter_Coupling} which concluded that $f_3=0$.

    \subsubsection{Anti-symmetric ansatz}

    We now proceed as in Ref.~\cite{auth:Takahashi_Matter_Coupling} and fix $c_{01}=0$ via a Lorentz transformation. In this case, by inspection of Eq.~\eqref{eq:Malphasecondderiv}, the only way to seemingly remove the higher order time derivative of $\cal X_\mu$ is to set $c_{11}=0$. This implies $f_3=0$. But, as anticipated, the anti-symmetric components of $\tensor{L}{^\mu_\nu}$ can be either removed by directly performing a Lorentz transformation of the tetrad base or by redefining the spinor field via
    \begin{align}
         \Psi \rightarrow  e^{ \frac{i}{2} T_{\mu\nu} \Sigma^{\mu\nu}} \Psi \,,
    \end{align}
    where $T_{\mu\nu}$ is an arbitrary anti-symmetric tensor. For a more detailed demonstration of this result, we refer the reader to Ref.~\cite{auth:us}. Therefore, instead of imposing directly $c_{11}=0$ one needs to check if the higher order time derivatives can be removed by performing a field redefinition. In the appendix \ref{app:Spinor_field_redefinition} we check for simplicity perturbatively up to order ${\cal O}(\mathbf{L}^2)$ that one recovers Eq.~\eqref{eq:M_mu_symmetric_ansatz}.

    \subsection{Arbitrary higher order spatial derivatives}
    Following \cite{Takahashi:2023vva} we can generalize the previous ansatz to include arbitrary higher order spatial derivatives. 
    Introducing the set of scalar quantities $\xi_i=\{X,{\cal Z}, {\cal X}^\mu \nabla_\mu {\cal Z}, ...   \}$ (see \cite{Takahashi:2023vva} for more details) we define
    \begin{align}
        S_{\mu} = \sum_i u_i h_\mu^\alpha \nabla_\alpha \xi_i\,,
    \end{align}
    where $u_i$ are arbitrary functions $u_i=u_i(\phi,\xi_j)$. 
    Using the ansatz
    \begin{align}
    L_{\mu\nu} = c_{00} \phi_{\mu} \phi_{\nu} + 2 c_{01} \phi_{(\mu} S_{\nu)} + c_{11} S_{\mu } S_\nu\,,
    \end{align}
    where $c_{IJ}=c_{IJ}(\phi,\xi_i)$, we obtain the same degeneracy condition as in the previous case by replacing ${\cal X}_\mu$ with $S_\mu$ leading to
\begin{align}\label{eq:c01higher}
    c_{01}^2 = - \frac{c_{11}^2 S_\mu S^\mu}{X}~\,.
\end{align}

The tetrad transformation corresponds to a disformal transformation of the metric given by
\begin{align}\label{eq:disformalSmu}
    \tilde g_{\mu\nu} = C^2 \left(  g_{\mu\nu} + f_1 \phi_\mu \phi_\nu + f_2 \phi_{(\mu} S_{\nu)} + f_3  S_\mu S_\nu \right)\,,
\end{align}
where the free function $f_i$ are given by Eqs. \eqref{eq:Definition_fi_metric} by replacing ${\cal Z}$ with $S_\mu S^\mu$. As before,
by inserting \eqref{eq:c01higher} into Eq.~\eqref{eq:Definition_fi_metric}, we again find that $f_3 = 2 C^2 c_{11}$. Note that, while we have only focused on the consistent matter coupling, it is left to derive the invertibility conditions for the disformal transformation \eqref{eq:disformalSmu}. Unfortunately, Ref.~\cite{Takahashi:2023vva} provides them for $f_3=0$. So, in order to obtain a class of generalized U-DHOST one needs to check if it is possible to obtain invertible disformal transformation with $f_3 \neq 0$ for a general $S_\mu$. This is out of the scope of this paper and we leave it for future work.

    \subsection{Disformal invariant couplings}

Another interesting case of disformal trasnformations with higher derivatives has been discussed in Ref.~\cite{Domenech:2023ryc}. There, one introduces a disformal invariant metric given by
\begin{align}
    \hat g_{\mu\nu}  = g_{\mu\nu} + \left( \frac{1}{X} +1 \right) \phi_\mu\phi_\nu\,,
\end{align}
and defines disformal invariant operators \cite{Domenech:2023ryc,Domenech:2025qny} via
\begin{align}
    {\cal G}_{\mu\nu} = \hat g_{\mu\nu}, \qquad {\cal N}^\mu = \hat g^{\mu\nu} \hat\nabla_\nu\phi, \qquad {\cal B}_{\mu\nu} = \hat \nabla_\mu \hat \nabla_\nu \phi, \qquad {\cal D}_{\mu\nu\alpha\beta} = \hat R_{\mu\nu\alpha\beta}\,,
\end{align}
where $\hat \nabla$ and $\hat R_{\mu\nu\alpha\beta}$ correspond to the covariant derivative and the Riemann tensor with respect to $\hat g$ respectively. Using these operators we can define an invertible disformal transformation via
\begin{align}
    \tilde g_{\mu\nu} = g_{\mu\nu} + D ( {\cal G}_{\mu\nu},{\cal N}^\mu,  {\cal B}_{\mu\nu}, {\cal D}_{\mu\nu\alpha\beta} ) \nabla_\mu \phi \nabla_\nu \phi\,.
\end{align}
We find that in this case the axial coupling $M_\alpha$ \eqref{eq:Definition_M_A} vanishes identically. Therefore, in order to construct a consistent matter coupling the conditions are the same as for bosons. It is straightforward to check that it is possible to construct combinations of the disformal operators which contain only first order time derivatives but higher order spatial derivatives in the unitary gauge (see \cite{Domenech:2023ryc}  for more details)  leading to another class of generalized U-DHOST.

\section{Higher derivative of a vector field}
\label{sec:single_vector}

In this last section, we focus on vector disformal transformation including derivatives of the vector field. Since calculations become involved, we discuss two simple, relevant examples, and leave a general analysis for future work. We note, though, that the class of allowed models is expected to be more restricted than in the scalar field case, as for instance we cannot impose the unitary gauge for the vector field as we do for the scalar field. We focus first on an example respective $U(1)$ gauge symmetry and then we move on to a more general case.

\subsection{$U(1)$ gauge symmetry}

Vector disformal transformations respecting a $U(1)$ gauge symmetry have been explored in \cite{DeFelice:2019hxb,Gumrukcuoglu:2019ebp}. In that case, the disformal transformation can only the depend on the field strength, that is $F_{\mu\nu}$, if we require parity conservation. The general transformation is then given by \cite{DeFelice:2019hxb}
\begin{align}\label{eq:gFF}
    \tilde g_{\mu\nu} = C^2 ( g_{\mu\nu} + B \tensor{F}{_\mu^\alpha} F_{\alpha\nu} )\,,
\end{align}
where $C$ and $B$ are functions of $\Tr \mathbf{F}^2$ and $\Tr \mathbf{F}^4$. In terms of the tetrads \eqref{eq:tetrads}, we have, in matrix notation, that
\begin{align}\label{eq:LF}
    \mathbf{L} =D \left( \mathbf{F}^2 - \frac{1}{4} \Trace \mathbf{F}^2 \mathds{1} \right)\,,
\end{align}
where $D=D(\Tr\mathbf{F}^2, \Tr \mathbf{F}^4)$.
From Eq.~\eqref{eq:Expression_P}, we obtain
\begin{align}\label{eq:PF}
    \mathbf{P} = -\frac{1}{1-a} \mathbf{L} + \frac{a}{1-a}  \mathds{1}\,,
\end{align}
where  we defined
\begin{align}
    a = D^2 \left( \frac{1}{4} \Tr\mathbf{F}^4 - \frac{1}{16} (\Tr\mathbf{F}^2)^2 \right)\,.
\end{align}

Using Eqs.~\eqref{eq:LF} and \eqref{eq:PF}, we find that the axial coupling \eqref{eq:Definition_M_A} is expressed in tensor components as
\begin{align}\label{eq:MAF}
    M_\alpha = &  \frac{1}{4} \tensor{\epsilon}{_\alpha^\beta^\gamma^\delta} \left( \delta^\nu_\delta + \tensor{P}{^\nu_\delta} \right) \tensor{P}{^\mu_\gamma} \nabla_\nu L_{\beta\mu} \nonumber \\
    =& - \frac{D^3}{4 (1-a)^2} \tensor{\epsilon}{_\alpha^\beta^\gamma^\delta} \left( \left( 1 + \frac{1}{4} \Tr\mathbf{F}^2 \right) \delta^{\nu}_\delta - \tensor{F}{^\nu_\sigma} \tensor{F}{^\sigma_\delta} \right) \tensor{F}{^\mu_\kappa} \tensor{F}{^\kappa_\gamma} \nabla_\nu (\tensor{F}{_\beta^\rho} \tensor{F}{_\rho_\mu})\,.
\end{align}
From the last term in Eq.~\eqref{eq:MAF}, we see that the axial coupling contains second time derivatives multiplied by first derivatives of the vector field . This means that the equations of motion contain third derivatives of the vector field. Namely, one find terms such as
\begin{align}
    \frac{\delta S}{\delta A^\mu} \supset &\frac{1}{2} J_5^\delta f \Big[ \tensor{\epsilon}{_\alpha^\gamma^\nu_\delta}   \tensor{F}{^\beta^\kappa} \tensor{F}{^\mu_\kappa}  + \tensor{\epsilon}{_\alpha^\gamma_\delta_\kappa}   \tensor{F}{^\beta^\mu} \tensor{F}{^\nu^\kappa} +  \tensor{\delta}{_\alpha^\mu} \tensor{\epsilon}{^\gamma_\delta_\kappa_\lambda}   \tensor{F}{^\beta^\kappa} \tensor{F}{^\nu^\lambda}  + \tensor{\epsilon}{^\gamma_\delta_\kappa_\lambda}   \tensor{F}{_\alpha^\kappa} \tensor{F}{^\mu^\lambda} g^{\beta\nu} \nonumber \\
    & +  \tensor{\epsilon}{^\gamma^\nu_\delta_\kappa} ( \tensor{F}{_\alpha^\mu} F^{\beta\kappa} - \tensor{F}{_\alpha^\kappa} F^{\beta\mu} ) - \tensor{\epsilon}{^\beta^\gamma^\nu_\delta} \tensor{F}{_\alpha^\kappa} \tensor{F}{^\mu_\kappa} + {\cal O}(F^4) \Big] \nabla_\beta \nabla_\gamma \tensor{F}{_\mu_\nu}\,.
\end{align}
These terms do not vanish in general. Thus, we conclude that the disformal coupling with $U(1)$ gauge symmetry \eqref{eq:gFF} is inconsistent with fermions. 

\subsection{General case}
If we do not impose $U(1)$ gauge symmetry, the analysis gets more involved as the number of possible terms is larger. For simplicity, let us excluding terms proportional to $\nabla_{(\mu} A_{\nu)}$. Then, a general vector disformal transformation reads
\begin{align}\label{eq:LgeneralA}
    L_{\mu\nu} = D_1 A_\mu A_\nu + D_2 A_{(\mu} F_{\nu)\alpha} A^\alpha + D_3 F_{\mu\alpha} \tensor{F}{^\alpha_\nu} + D_4 A_{(\nu}F_{\mu)\alpha} \tensor{F}{^\alpha_\beta} A^\beta + ...\,,
\end{align}
where the free functions depend on $A_\mu$ and $F_{\mu\nu}$, that is $D_i= D_i(A_\mu, F_{\mu\nu})$, and the dots denote higher order contractions between $A_\mu$ and $F_{\mu\nu}$.

As the full analysis is beyond the scope of this paper, let us consider simple yet relevant example where we fix $D_1=D_1(A_\mu, F_{\mu\nu})\neq0$ and $D_{i>1} =0$. In that case the axial coupling \eqref{eq:Definition_M_A} reads
\begin{align}
    M_\mu = - \frac{ D_1^2 X }{8  (1 + D_1 X )} \tensor{\epsilon}{_\mu^\beta^\gamma^\delta} A_\delta F_{\beta\gamma}\,,
\end{align}
where $X=A_\mu A^\mu$. Since the coupling to the axial current does involve derivatives of $D_1$, we conclude that it has a consistent matter coupling despite $D_1$ depending on $F_{\mu\nu}$.

It is interesting to note that, if we  performed the inverse transformation to the Jordan frame, the gravity sector would depend on covariant derivatives of the field strength tensor. This would normally signal the presence of ghost degrees of freedom. For instance, such terms are not present in generalized Proca theories \cite{Heisenberg:2014rta} or generalizations of it \cite{Kimura:2016rzw} as they contain at maximum first derivatives of the vector field. However, we have shown that the corresponding metric disformal transformation to \eqref{eq:LgeneralA}, which is given by
\begin{align}
    \tilde g_{\mu\nu} = C^2 \left( g_{\mu\nu} + D_1 \left( 2 + D_1 X \right) A_\mu A_\nu \right)~\,,
\end{align}
has a consistent matter coupling. We expect that such disformal transformation leads to a new class of generalized vector theories with higher order derivatives but without ghost degrees of freedom, as long as the transformation is invertible. We plan to study such generalized higher derivative vector theories in the future.

\section{Conclusion}
\label{sec:Conclusion}

We studied the impact of a disformal coupling to the fermionic sector, which leads to derivative couplings between the disformal fields and the fermionic axial current in general. We demonstrated that such a coupling may lead to higher derivative terms in the equations of motion and, therefore, to Ostrogradsky ghost instabilities. Although our formalism can be applied to general cases, we have focused for concreteness on the cases of multi-scalar field and higher-derivative scalar and vector fields disformal couplings.

Already for disformal transformation consisting of two scalar fields we demonstrate that one needs to impose degeneracy conditions on the form of the field space metric of the disformal transformation. The degeneracy conditions leads to a degenerate field space metric such that effectively the disformal transformation only points along one direction leading to an identically vanishing coupling to the axial current. This demonstrates that in contrast to single scalar field disformal transformation the form of the transformation is highly restricted. We conjecture that similar statements apply for $N$ scalar fields such that the only valid disformal transformation points effectively along one direction.
The new class of disformal transformation may be helpful in the exploration of the recently proposed two-field Horndeski model \cite{Horndeski:2024hee} or for the construction of a corresponding two-field DHOST model, which we leave for future work. Interestingly, we have shown that a term previously postulated in Ref.~\cite{Ohashi:2015fma}  naturally appears from the two-scalar field disformal transformation of the Einstein-Hilbert action.

The formalism can also be applied to the single field disformal transformation with higher derivatives. In doing so, we extended the conditions derived in Ref.~ \cite{auth:Takahashi_Matter_Coupling}, leading to a larger class of disformal Horndeski theories. 
Similarly, one can construct a viable disformal transformation of a single vector field depending directly on the field strength, which can be used to explore new classes of Proca theories beyond current approaches, such as those in Refs.~ \cite{Heisenberg:2016eld,Kimura:2016rzw}. We leave it for future work.

\begin{acknowledgments}
We would like to thank Kazufumi Takahashi for the helpful comments and discussions. 
This research is supported by the DFG under the Emmy-Noether program, project number 496592360, and by the JSPS KAKENHI grant No. JP24K00624.
\end{acknowledgments}

\appendix

    \section{Transformation of the spin connection and Dirac Lagrangian}
    \label{sec:Transformation_Tensor_Components}

        Here we review the disformal transformation of the spin connection and the Dirac Lagrangian using tensor components, for more details see Ref.~\cite{auth:us}. 
        Employing the transformation rule of the individual tetrads and that of the connection, we get
            \begin{align}
                \tensor{\tilde{\omega}}{_A_B_C} & = \tensor{\tilde{e}}{_C^\mu}\left(\tilde{\omega}_{AB}\right)_\mu= \tensor{\tilde{e}}{_C^\mu}\tensor{\tilde{e}}{_{[A}^\nu}\tilde{\nabla}_\mu \tensor{\tilde{e}}{_{B]}_\nu}
                = \tensor{\tilde{e}}{^\mu_C}\tensor{\tilde{e}}{_{[A}^\nu}\nabla_\mu \tensor{\tilde{e}}{_{B]}_\nu} - \tensor{\tilde{e}}{_C^\mu}\tensor{\tilde{e}}{_{[A}^\nu}\tensor{\tilde{e}}{_{B]}_\alpha}\tensor{\mathcal{K}}{^\alpha_\mu_\nu}\,,\label{eq:App_Connection_Coeff}
            \end{align}
        where
            \begin{equation}
                \tensor{\mathcal{K}}{^\alpha_\mu_\nu}=\frac{1}{2}\tilde g^{\alpha\lambda}\left(\nabla_\mu \tilde g_{\nu\lambda}+\nabla_\nu \tilde g_{\mu\lambda}-\nabla_\lambda \tilde g_{\mu\nu}\right)\,.
                \label{eq:K_tensor}
            \end{equation}
        Expressing eq.~\eqref{eq:K_tensor} in terms of the tetrads and inserting it into~\eqref{eq:App_Connection_Coeff} we obtain that\footnote{One can show that
        \begin{equation}
                \tensor{\tilde{e}}{_A^\nu}\nabla_\mu \tensor{\tilde{e}}{_B_\nu} = \tensor{e}{_A^\nu}\nabla_\mu \tensor{e}{_B_\nu} + \tensor{\eta}{_A_B}\nabla_\mu \ln C + \tensor{e}{_A^\alpha}\tensor{e}{_B_\beta}\tensor{(\mathcal{U}^{-1})}{^\nu_\alpha}\nabla_\mu\tensor{\mathcal{U}}{^\beta_\nu}\,.
            \end{equation}}
        \begin{align}\label{eq:Conn_Coef_Spacetime}
                \tensor{\tilde{\omega}}{_A_B_C} &=
    \tilde e^\mu_C \tilde  e^\nu_{[A}  \nabla_\mu \tilde e_{\nu B]}-\tilde  e^\nu_{C} \tilde e^\mu_{[A}  \nabla_\mu \tilde e_{\nu B]}-\tilde e^\mu_{[A} \tilde  e^\nu_{B]}  \nabla_\mu \tilde e_{\nu C}\nonumber\\&
    =\tensor{\tilde{e}}{_C^\mu}\tensor{e}{_{[A}^\nu}\nabla_\mu \tensor{e}{_{B]}_\nu} - 2\tensor{\tilde{e}}{_{[A}^\mu}\tensor{\eta}{_{B]}_C}\nabla_\mu\ln C + \left( \tensor{e}{_C^\rho}\tensor{e}{_{[A}^\alpha}\tensor{e}{_{B]}_\beta} - \tensor{e}{_{[A}^\rho}\tensor{e}{_C^\alpha}\tensor{e}{_{B]}_\beta} - \tensor{e}{_{[A}^\rho}\tensor{e}{_{B]}^\alpha}\tensor{e}{_C_\beta}  \right)\tensor{\hat{\mathcal{S}}}{^\beta_\rho_\alpha}\,,
            \end{align}
        where we defined
            \begin{equation}\label{eq:S_Spacetime_Basis}
                \tensor{\hat{\mathcal{S}}}{^\beta_\rho_\alpha} = \frac{1}{C}\tensor{(\mathcal{U}^{-1})}{^\mu_\rho}\tensor{(\mathcal{U}^{-1})}{^\nu_\alpha}\nabla_\mu \tensor{\mathcal{U}}{^\beta_\nu}\,.
            \end{equation}

        Note that the spin connection \eqref{eq:Conn_Coef_Spacetime} has a totally antisymmetric component, namely
         \begin{equation}\label{eq:Conn_Coef_Spacetime2}
                \epsilon^{ABCD}\left(\tensor{\tilde{\omega}}{_A_B_C}-e^\alpha_C\tensor{(\mathcal{U}^{-1})}{^\mu_\alpha}\tensor{\tilde{\omega}}{_A_B_\mu}\right) = T^D\,,
            \end{equation}
            where we introduced
            \begin{align}
             T^D=\epsilon^{DEFG}\tensor{e}{_{E}^\rho}\tensor{e}{_F^\alpha}\tensor{e}{_{G}_\beta}\tensor{\hat{\mathcal{S}}}{^\beta_\rho_\alpha}=-\frac{4}{C}e^D_\alpha M^\alpha\,,
            \end{align}
            and in the last step we used $M^\alpha$ as given in Eq.~\eqref{eq:Definition_M_A}.
            
        To obtain the action \eqref{eq:Transformed_Dirac_Lagrangian}, we use that
        \begin{align}
        \label{eq:two-sided derivative}
            \tilde{\bar \Psi} \gamma^A \tilde{\overset{\leftrightarrow}{\nabla}}_A \tilde \Psi = - \partial_A \tilde{\bar \Psi} \gamma^A \tilde \Psi + \tilde{\bar \Psi } \gamma^A \partial_A \tilde \Psi + \frac{i}{4}  \tilde{\bar \Psi} \{\gamma^A,\Sigma^{BC}\} \tensor{\tilde \omega}{_B_C_A} \tilde \Psi\,,
        \end{align}
        in addition to the following identity of the gamma matrices, that is
        \begin{align}
        \label{eq:identitysigma}
            \{\gamma^A, \Sigma^{BC} \} = -\tensor{\epsilon}{_D^A^B^C}  \gamma^D \gamma^5 \,.
        \end{align}
        The terms which are not totally antisymmetric in \eqref{eq:Conn_Coef_Spacetime2}, that is those except for $T^D$, are absorbed in a rescaling the spinor field given by \eqref{eq:Spinor_rescaling}.
        It is interesting to note that the fermions in the hermitian formulation of the Dirac Lagrangian only couple to the total antisymmetric components of the spin-connection, besides the overall modification of the kinetic term.

\section{Spinor field redefinition}
\label{app:Spinor_field_redefinition}

In this appendix we provide details on how to remove any antisymmetric component of the tetrad disformal transformation via a spinor field redefintion. As discussed in Ref.~\cite{auth:us}, one can do a general  spinor field redefinition via
\begin{align}
    \Psi \rightarrow e^{ \frac{i}{2} T_{\alpha\beta}   \Sigma^{\alpha\beta}} \Psi\,,
\end{align}
where $\mathbf{T}$ is an arbitrary anti-symmetric matrix. Next, we use the identity given by
\begin{align}
    e^{- \frac{i}{2} T_{\mu\nu} \Sigma^{\mu\nu} } \gamma^\alpha e^{\frac{i}{2} T_{\mu\nu} \Sigma^{\mu\nu}} = \tensor{\Lambda}{^\alpha_\beta} \gamma^\beta \simeq \left( \tensor{\delta}{^\alpha_\beta} + \tensor{T}{^\alpha_\beta} \right) \gamma^\beta\,,
\end{align}
where in the last step we expanded at leading order in $\mathbf{T}$.  By using such field redefinition in Eq. \eqref{eq:Transformed_Dirac_Lagrangian} we obtain that
\begin{align}
    S_{\rm Dirac} = -\int \md^4x\, \det e\, & \Big[ \frac{1}{2} \tensor{({\cal U}^{-1})}{^\mu_\nu} \Big( \bar \Psi  e^{- X} \gamma^\nu \nabla_\mu \left(  e^{ X} \Psi \right) - \nabla_\mu \left( \bar \Psi  e^{- X } \right) \gamma^\nu  e^{ X }  \Psi  \Big) \nonumber \\
    & + i M_\mu \bar \Psi   e^{- X } \gamma^\mu \gamma^5  e^{ X} \Psi + C \bar \Psi m \Psi \Big] \,, 
\end{align}
where we have introduced the notation $ X = \frac{1}{2} T_{\alpha\beta} \Sigma^{\alpha\beta}$.
Using the relation
\begin{align}
    \nabla e^X = \int_0^1 \md \alpha e^{\alpha X} \nabla X e^{(1-\alpha) X}
\end{align}
we obtain after straightforward calculations
\begin{align}
    S_{\rm Dirac} = - \int \md^4x\, \det e\, & \Big[  \frac{1}{2} \tensor{({\cal U}^{-1})}{^\mu_\nu} \tensor{\Lambda}{^\nu_\beta} (\bar \Psi \gamma^\beta \overset{\leftrightarrow}{\nabla}_\mu \Psi )  +  \frac{1}{2} \tensor{({\cal U})}{^\mu_\nu} \bar \Psi  \Big\{ \tensor{\Lambda}{^\nu_\beta} \gamma^\beta, \int \md \alpha e^{-\alpha X } \nabla_\mu X e^{\alpha X}  \Big\}  \Psi \nonumber \\
    & + i M_\mu \tensor{\Lambda}{^\mu_\nu} J_5^\nu + C \bar \Psi m \Psi \Big]~. \label{eq:Full_Lorentz_transformation}
\end{align}
Note that if we expand the kinetic term at leading order in $\mathbf{L}$ and in ${\mathbf{T}}$, namely
\begin{align}
 \tensor{({\cal U}^{-1})}{^\mu_\nu} \tensor{\Lambda}{^\nu_\beta} \bar \Psi \gamma^\beta \overset{\leftrightarrow}{\nabla}_\mu \Psi \simeq \left( \tensor{\delta}{^\mu_\nu} - \tensor{L}{^\mu_\nu} + \tensor{T}{^\mu_\nu} \right) \bar \Psi \gamma^\nu \overset{\leftrightarrow}{\nabla}_\mu \Psi
\end{align}
we can remove the antisymmetric component contractions at leading order by defining $T_{\mu\nu} = L_{[\mu\nu]}$. One may also do it at all orders in ${\mathbf{L}}$ by iteratively finding the appropriate ${\mathbf{T}}$. We will not pursue it here as the explicit result is not particularly illuminating.
Besides the kinetic coupling, which is now explicitly symmetric, there remain couplings to the axial current. 
In the following we only focus on the leading order terms and expand the integral up to second order in $\mathbf{T}^2$. Using that
\begin{align}
    [ \Sigma^{\mu\nu}, \Sigma^{\rho\lambda}] = g^{\mu\lambda} \Sigma^{\nu\rho} + g^{\nu\rho} \Sigma^{\mu\lambda} - g^{\mu\rho} \Sigma^{\nu\lambda} - g^{\nu \lambda} \Sigma^{\mu\rho} 
\end{align}
and Eq. \eqref{eq:identitysigma} we can note that the second term in Eq. \eqref{eq:Full_Lorentz_transformation} contributes to the coupling to the axial current. Up to quadratic order in $\mathbf{L}^2$ using $T_{\mu\nu} = L_{[\mu\nu]}$ we get for the axial coupling
\begin{align}
    S_{\rm Dirac} \supset - \int \md^4x\, \det e\, & \Big[  \frac{1}{2} \tensor{({\cal U}^{-1})}{^\mu_\nu} \bar \Psi  \Big\{ \tensor{\Lambda}{^\nu_\beta} \gamma^\beta, \int \md \alpha e^{-\alpha X } \nabla_\mu X e^{\alpha X}  \Big\}  \Psi + i M_\mu \tensor{\Lambda}{^\mu_\nu} J_5^\nu \Big] \nonumber \\
    \simeq   - \int \md^4x\, \det e\, & \Big[  \frac{1}{2} \tensor{({\cal U}^{-1})}{^\mu_\nu} \tensor{\Lambda}{^\nu_\beta} \bar  \Psi \{\gamma^\beta, \nabla_\mu X - \frac{1}{2} [X, \nabla_\mu X] \} + i M_\mu \tensor{\Lambda}{^\mu_\nu} J_5^\nu \Big] \nonumber \\
    \simeq  - i \int \md^4x\, \det e\, & \Big[  \frac{1}{4} \tensor{({\cal U}^{-1})}{^\mu_\nu}  \tensor{\Lambda}{^\nu_\beta} \nabla_{\mu} T_{\rho\lambda}  \left(  \{ \gamma^\beta, \Sigma^{\rho\lambda} \}  - \frac{1}{2}  T_{\sigma\xi} \{ \gamma^\beta, [\Sigma^{\sigma\xi},\Sigma^{\rho\lambda} ]\} \} \right) +  M_\mu \tensor{\Lambda}{^\mu_\nu} J_5^\nu\Big] \nonumber \\
     \simeq   - i \int \md^4x\, \det e\, & \frac{1}{4}  \Big[ \tensor{\epsilon}{_\alpha^\beta^\rho^\lambda} \Big( - \nabla_\lambda T_{\beta\rho} + \nabla_\lambda L_{\beta\rho} -\tensor{T}{^\mu_\beta} \nabla_\mu T_{\rho\lambda} - 2 \tensor{T}{^\gamma_\lambda} \nabla_\beta T_{\gamma\rho} - \tensor{L}{^\mu_\beta} \nabla_\mu T_{\rho\lambda} \nonumber \\
     &  -  \tensor{L}{^\mu_\lambda} \nabla_\mu  L_{\beta\rho} - \tensor{L}{^\nu_\rho} \nabla_\lambda L_{\beta\nu} \Big) + \tensor{\epsilon}{_\sigma^\beta^\rho^\lambda} \tensor{T}{^\sigma_\alpha}  \nabla_\lambda L_{\beta\rho} \Big]  J_5^\alpha
\end{align}
First, we can note that the linear order terms exactly cancel each other due to the contraction with the Levi-Civita tensor. This is expected as the term is only present for an anti-symmetric disformal matrix which can be removed by the Lorentz transformation of the tetrads. 

Our last step is to recover the results of Sec.~\ref{sec:secondderivativesscalar}. To do so, we apply the formalism to the antisymmetric disformal transformation in Eq. \eqref{eq:Lmunusecondder} with $c_{01}=0$ to relate $\tensor{ T}{_\mu_\nu}$ with the scalar field derivatives, namely
\begin{align}
    \tensor{ T}{_\mu_\nu} = c_{10} {\cal X}_{[\mu} \phi_{\nu]}~\,.
\end{align}
Then, we obtain, up to quadratic order in $\mathbf{L}$, for the higher order time derivative terms in the Dirac action that
\begin{align}
    S_{\rm Dirac} \supsetsim  i \int \md^4x\, \det\,e\, \Big[ \left( \frac{c_{10}^2}{8} + \frac{c_{11}^2 {\cal Z}}{4 X} \right)  \tensor{\epsilon}{_\delta^\alpha^\beta^\gamma} \phi_\gamma {\cal X}_\beta \phi^\mu \nabla_\mu {\cal X}_\alpha + {\cal O}(\mathbf{L}^3)  \Big]~.
\end{align}
Rescaling the parameter $c_{10} \rightarrow \sqrt{2} c_{10}$ we recover the result in Eq. \eqref{eq:M_mu_symmetric_ansatz} up to quadratic order in ${\cal O}(\mathbf{L}^2)$. Note, that we have to rescale the solution to account that setting $c_{01}=0$ and then symmetrize the disformal matrix is not equivalent to directly symmetrize the original ansatz.

\if{}
    
\section{General case}
The problematic terms spoiling the degeneracy condition leading to a potential ghost degree of freedom are solely coming from the time derivative of ${\cal U}$. Therefore, splitting the metric into a spacelike and time projection
\begin{align}
    \delta^A_B = h^A_B - n^A n_B
\end{align}
we will introduce
\begin{align}
    \tensor{{\hat P}}{^A_B} = h^A_K h_B^L \tensor{P}{^K_L}, \qquad \tensor{P}{_n_L} = n_A \tensor{P}{^A_L}, \qquad \hat P_{nL} = n^A h_L^B P_{AB}, \qquad P_{nn} = n^A n^B P_{AB} 
\end{align}
and similarly for $L$ and $\nabla_n = n^A \nabla_A$. Using it, we get 
\begin{align}
    M_A \supset & \frac{1}{4}  \tensor{\epsilon}{_A^B^C^D} \left( - n_D \tensor{P}{^T_C} \nabla_n L_{BT} - P_{nD} \tensor{P}{^T_C} \nabla_n L_{BT} - n_D \nabla_n L_{BC} \right) \nonumber \\
    \supset & \frac{1}{4} \tensor{\epsilon}{_A^B^C^D} \Big[ \left( -n_D ( \tensor{\hat P}{^T_C} -\hat P_{nC} n^T ) -  (\hat P_{nD} - n_D P_{nn})  \tensor{P}{^T_K} h^K_C + n_C \tensor{P}{^T_n} \hat P_{nD}   \right) \nabla_n L_{BT}  - n_D \nabla_n L_{BC} \Big] \nonumber \\
    \supset & \frac{1}{4}\tensor{\epsilon}{_A^B^C^D} \Big[ n_D \left(- \tensor{\hat P}{^T_C} + \hat P_{nC} n^T - \hat P_{nC} (\tensor{\hat P}{^T_n} - P_{nn} n^T)  + P_{nn} (\tensor{\hat P}{^T_C} - \hat P_{nC} n^T)\right) \nabla_n L_{BT} \nonumber \\
    & - \hat P_{nD} (\tensor{\hat P}{^T_C} - \hat P_{nC} n^T) \nabla_n L_{BT} - n_D \nabla_n L_{BC} \Big] \nonumber \\
    \supset & \frac{1}{4}\tensor{\epsilon}{_A^B^C^D} \Big[ n_D \left(- \tensor{\hat P}{^T_C} + \hat P_{nC} n^T - \hat P_{nC} \tensor{\hat P}{^T_n}   + P_{nn} \tensor{\hat P}{^T_C} )\right) \nabla_n L_{BT} - \hat P_{nD} (\tensor{\hat P}{^T_C} - \hat P_{nC} n^T) \nabla_n L_{BT} - n_D \nabla_n L_{BC} \Big] 
\end{align}
Note, that I do not assume that $L$ is symmetric at the moment. Furhter, we can use that
\begin{align}
    \nabla_\mu n_\nu = K_{\mu\nu} - n_\mu a_\nu
\end{align}
where $a_\mu$ is the acceleration of the normal vector and $K_{\mu\nu}$ the extrinsic curvature. Therefore
\begin{align}
    \nabla_n L_{BC} = & \nabla_n \left( \hat L_{KL} h^K_B h^L_C - \hat L_{nK} n_B h^K_C - \hat L_{Kn} n_C h^K_B +  L_{nn} n_B n_C   \right) \nonumber \\
    =& h^K_B h^L_C \nabla_n \hat L_{KL} - a^K n_B \hat L_{KC}  -  a^L n_C \hat L_{BL} - n_B h^K_C \nabla_n \hat L_{nK} - \hat L_{nC} a_B + \hat L_{nK} a^K n_C n_B \nonumber \\
    & - n_C h^K_B  \nabla_n \hat L_{Kn} - a_C \hat L_{Bn} + n_C \hat L_{Kn} a^K n_B + n_B n_C \nabla_n L_{nn} + ( n_B a_C + n_C a_B) L_{nn}
\end{align}
Note, that the acceleration does not contain time derivatives so that we can ignore it for our purposes. 
Finally, we get
\begin{align}
    M_A \supset & \frac{1}{4}\tensor{\epsilon}{_A^B^C^D} \Big[ n_D \left( - \tensor{\hat P}{^T_C}  - \hat P_{nC} \tensor{\hat P}{^T_n}   + P_{nn} \tensor{\hat P}{^T_C}  \right) h^K_B h^L_T \nabla_n \hat L_{KL}  + n_D \hat P_{nC} h^K_B \nabla_n \hat L_{Kn}  \nonumber \\
    & - \hat P_{nD} \tensor{\hat P}{^T_C} ( h^K_B h^L_T \nabla_n \hat L_{KL} - n_B h^K_T \nabla_n \hat L_{nK}  )  - n_D h^K_B h^L_C \nabla_n \hat L_{KL} \Big] \nonumber \\
    \supset & \frac{1}{4}\tensor{\epsilon}{_A^B^C^D}  \Big[ \left( n_D h^K_B h^R_C\left( - \tensor{\hat P}{^L_R}  - \hat P_{nR} \tensor{\hat P}{^L_n}   + P_{nn} \tensor{\hat P}{^L_R}   \right) - \hat P_{nD} \tensor{\hat P}{^L_C} h^K_B  - n_D h^K_B h^L_C \right)    \nabla_n \hat L_{KL} \nonumber \\
    & +  n_D \hat P_{nC} h^K_B  \nabla_n \hat L_{Kn} + n_D \hat P_{nC} \tensor{\hat P}{^K_B} \nabla_n \hat L_{nK}  \Big]
\end{align}
We can already make some first observations: First, any tensor having only time-time components will not lead to any problematic couplings. 
Second, if we assume that $\vert P \vert  \ll 1$ we need to require that $\hat L_{AB}$ is symmetric.

\section{Easy consistent couplings}

Let us demonstrate some easy consistent couplings. 

\subsection{Conformal couplings}
Trivially, any conformal transformation leads to $M_A=0$. Note that such transormation with higher derivatives in the free function $C$ have been discussed in ...

\subsection{Purely single scalar disformal transformation}
Let us assume that the disformal transformation can be expressed as
\begin{align}
    L_{AB} = D \nabla_A Z \nabla_B Z
\end{align}
where $Z$ can be any scalar quantity and $D$ any function of spacetime. In this case, $M_A =0$ by symmetry. A priori, the transformation might not always be invertible. However, besides the standard disformal transformation there are two interesting subcases which have been discusses in the literature which are invertible. 

\textbf{Disformal invariant couplings:}
\begin{align}
    \tilde g_{\mu\nu} = g_{\mu\nu} + D ( {\cal D}_{\mu\nu\alpha\beta}, {\cal B}_{\mu\nu}, {\cal G}_{\mu\nu} ) \nabla_\mu \phi \nabla_\nu \phi
\end{align}
where the free function depend on disformal invariant scalars. These can contain higher spatial derivatives, i.e. ${\cal D} \simeq R[h]$ in the unitary gauge.

\subsection{Antisymmetric structure}
Another easy example is given by 
\begin{align}
    L_{AB} = h^L_A n_B n^K T_{LK}
\end{align}
It is straightforward to check that $P$ has the same structure as $L$. 
The transformation has only 4 independent components. Such a structure has been used in ... to construct disformal transformation with arbitrary higher spatial derivatives of the scalar field, i.e.
\begin{align}
    L_{\mu\nu} = c_{0} \phi_\mu \phi_\nu + \phi_\nu \sum_i c_i h^\alpha_\mu \nabla_\alpha \chi_i
\end{align}
where $\chi_i = \{X, {\cal Z}, {\cal U}, {\cal V},... \}$ (see 2307.08814)

\subsection{Diagonal spatial matrix}

Next, let us consider the case that the spatial part of the matrix $P$ and $L$ are diagonal $\tensor{L}{^K_L} = \hat L h^K_{L}$ and similarly for $P$. In this case we can directly obtain
\begin{align}
   M_A \supset & \frac{1}{4}\tensor{\epsilon}{_A^B^C^D}  \Big[ \left( n_D h^L_B h^R_C\left( - \tensor{\hat P}{^L_R}  - \hat P_{nR} \tensor{\hat P}{^L_n}   + P_{nn} \tensor{\hat P}{^L_R}   \right) - \hat P_{nD} \tensor{\hat P}{_B_C}   - n_D h_{BC} \right)    \nabla_n \hat L \nonumber \\
    & +  n_D \hat P_{nC} h^K_B  \nabla_n \hat L_{Kn} + n_D \hat P_{nC} \tensor{\hat P}{^K_B} \nabla_n \hat L_{nK}  \Big] \nonumber \\
    =& \frac{1}{4}\tensor{\epsilon}{_A^B^C^D} \Big[ - n_D \hat P_{nC} \hat P_{Bn} \nabla_n \hat L + n_D \hat P_{nC} h^K_B \nabla_n \hat L_{Kn} + n_D \hat P_{nC} \hat P \nabla_n \hat L_{nB} \Big]
\end{align}
If we assume that $\hat P_{nC}=0$ as before all the terms vanish and we can see that we can generalize the previous approach by adding a diagonal spatial matrix. On the other hand, for a symmetric matrix we get the constraints 
\begin{align}
    M_A \supset \frac{1}{4}\tensor{\epsilon}{_A^B^C^D} n_D (1 + \hat P) \hat P_{nC} \nabla_n \hat L_{Bn}
\end{align}
Note, that a diagonal spatial matrix is redundant in the sense that we can absorb it into the conformal factor by shifting the $L_{nn}$ component, which does not lead to any issues wrt the consistency conditions on the matter coupling.

\subsection{Higher order spatial derivatives}
\subsubsection{Two-fields}

Let us review the case with
\begin{align}
    L_{\mu\nu} = c_{00} \phi_\mu \phi_\nu + 2 c_{01} \phi_{(\mu} {\cal X}_{\nu)} + c_{11} {\cal X}_{\mu} {\cal X}_\nu
\end{align}
where ${\cal X}_\mu = h_\mu^\nu \nabla_\nu X$.
The potential dangerous terms are coming from
\begin{align}
        M_A \supset &  \frac{1}{4} \tensor{\epsilon}{_A^B^C^D} \left(  \tensor{c}{_0_1} \tensor{d}{_0_1} + \frac{{\cal Z}}{X} c_{11} d_{11} + (d_{00} d_{11} -d_{01}^2 ) {\cal Z} c_{11} \right) {\cal X}_C \phi_D  \phi^Q \nabla_Q {\cal X}_B \nonumber \\
        = & - \frac{1}{4} \frac{c_{01}^2 X + c_{11}^2 {\cal Z}}{X T} \tensor{\epsilon}{_A^B^C^D} {\cal X}_C \phi_D  \phi^Q \nabla_Q {\cal X}_B
    \end{align}
    where in the second step we have used that the functions $d_{ij}$ and $T$ are given by
    \begin{align}
        d_{00} = & - \frac{c_{00} + c_{00} c_{11} {\cal Z} - c_{01}^2 {\cal Z} }{T}, \\
        d_{01} =& - \frac{c_{01}}{T}, \\
        d_{11} =& - \frac{c_{11} + c_{00} c_{11} X - c_{01}^2 X }{T}, \\
        T= & 1+ c_{00} X + c_{11} {\cal Z} + X {\cal Z} (c_{00} c_{11} - c_{01}^2)
    \end{align}
    In order to remove the dangerous term we can impose the degeneracy condition
    \begin{align}
        c_{01}^2 = - \frac{c_{11}^2 {\cal Z}}{X}
    \end{align}
    Therefore, it is possible to find a disformal transformation with $c_{11} \neq 0$ and consequently $f_3\neq 0$.

    Note, the important difference to the case in the introduction. In this case, we always assume $\phi_\mu \propto n_\mu$. It is straightforward to check that under this assumption there are no higher order time derivatives in $M_A$ for the two scalar fields even without imposing any degeneracy condition, only higher order spatial derivatives.

\subsection{Bigravity}

It may be interesting to also include a short discussion about disformal transformation with a second metric. This has been extensively discussed in context of bigravity for bosons which already breaks the degeneracy condition and revives the Boulware-Deser ghost. Up to my knowledge noone has discussed it in the context of fermions. 

\fi

\bibliography{refs.bib}

\end{document}